\documentclass[11pt]{article}

\usepackage[final]{graphicx}
\usepackage{float}
\usepackage{subcaption}
\usepackage{amsmath}
 \usepackage{qcircuit}
 \usepackage{authblk}
 \usepackage{url}
 \newtheorem{definition}{Definition}
 \newtheorem{example}{Example}
\usepackage[ruled,vlined]{algorithm2e}
\newtheorem{Rule}{Simplification Rule}

\SetKwRepeat{Do}{do}{while}  
 \usepackage{algpseudocode}
 
\usepackage[belowskip=-10pt,aboveskip=10pt]{caption}

\title{Multi-strategy Based Quantum Cost Reduction of Quantum Boolean Circuits}
\author[1]{Taghreed Ahmed \thanks{taghreed\_ebed@yahoo.com}}
\author[1,2]{Ahmed Younes \thanks{ayounes@alexu.edu.eg}}
\author[1,3]{Islam Elkabani \thanks{islam.Kabani@alexu.edu.eg}}
\affil[1]{ Department of Mathematics and Computer Science, Faculty of Science, Alexandria University, Alexandria, Egypt }
\affil[2]{Alexandria Quantum Computing Group, Alexandria University, Alexandria, Egypt}
\affil[3]{Faculty of Computer Science and Engineering, Alamein International University, El-Alamein, Egypt}

\date{\today}
\begin{document}
\maketitle
\begin{abstract}
The construction of quantum computers is based on the synthesis of low-cost quantum circuits. The quantum circuit of any Boolean function expressed in a Positive Polarity Reed-Muller $PPRM$ expansion can be synthesized using Multiple-Control Toffoli ($MCT$) gates. This paper proposes two algorithms to construct a quantum circuit for any Boolean function expressed in a Positive Polarity Reed-Muller $PPRM$ expansion. The Boolean function can be expressed with various algebraic forms, so there are different quantum circuits can be synthesized for the Boolean function based on its algebraic form. The proposed algorithms aim to map the $MCT$ gates into the $NCV$ gates for any quantum circuit by generating a simple algebraic form for the Boolean function. The first algorithm generates a special algebraic form for any Boolean function by rearrangement of terms of the Boolean function according to a predefined degree of term $d_{term}$, then synthesizes the corresponding quantum circuit. The second algorithm applies the decomposition methods to decompose $MCT$ circuit into its elementary gates followed by applying a set of simplification rules to simplify and optimize the synthesized quantum circuit. The proposed  algorithms achieve a reduction in the quantum cost of synthesized quantum circuits when compared with relevant work in the literature. The proposed algorithms synthesize quantum circuits that can applied on IBM quantum computer.
\end{abstract}


\section{Introduction}
\label{intro}
The  analysis and design of Boolean functions are important in engineering and science because they have several applications in electrical engineering, game theory, logic and cryptography \cite{17,book}. Synthesis of quantum circuits for Boolean functions has taken attention from researchers due to quantum mechanical properties of quantum circuits \cite{16}. There are many applications for quantum circuits in various technologies such as optical computing \cite{19}, Quantum Dot Cellular Automata \cite{4,6}, low-power CMOS \cite{5}, DNA computing \cite{3}, nanotechnology \cite{19,23}, quantum computation \cite{doi:10.1080/17445760500355678} and quantum image processing \cite{10}. One of objectives of quantum circuit design is to minimize the quantum cost and the number of gates to get a quantum circuit with minimal quantum volume which can run efficiently on a quantum computer with less quantum runtime \cite{19}. There are many algorithms and methods to design reversible and quantum circuits to minimize quantum cost. One of these methods constructs the reversible circuits with extra ancillary bits that will increase the hardware cost \cite{21}. A rule-based optimization algorithm \cite{Arabzadeh:2010:ROR:1899721.1899916} uses multiple control Toffoli gates with both negative and positive controls which merges Toffoli gates with common target and introduces rules for optimizing reversible circuits and sub-circuits that have a common target. Another method \cite{12} uses a Quantum Operator Form $(QOF)$ to minimize the quantum cost that allows the usage of different quantum gates $CNOT$, $Controlled-V$ and $Controlled-V\dagger$. In \cite{11}, the Multi Controls Target $MCT$ circuit is mapped into Linear Nearest Neighbor $LNN$ circuit by decomposing $MCT$ gates with decomposition methods then synthesizing a nearest neighbor quantum circuit and simplifying the circuit to minimize the quantum cost of reversible circuit. An exact method \cite{22} is proposed to synthesize a nearest neighbor compliant quantum circuit for a given reversible function. The exact method \cite{22} applies exhaustive search to get a specific qubit placement to synthesize a quantum circuit with minimal quantum cost. A reorder algorithm given in \cite{journals/qip/AhmedYE18} synthesizes the reversible circuits for Boolean functions with $n$ variables represented as a Positive Polarity Reed-Muller $PPRM$ expansion. The main idea of the reorder algorithm is to rewrite terms of the Boolean function after applying factorization algorithm. Another method \cite{7} based on genetic algorithms to get fixed and mixed polarity Reed-Muller expansions and optimize the circuit realization of a Boolean function. In \cite{25}, a method proposes a reversible circuit combining factorization and Boolean expression diagram (BED) to synthesize a quantum circuit which takes cubes in the exclusive-sum-of-products (ESOP) expansion of a Boolean function as input represented by decision graph representation. These cubes are factorized by leveraging algebraic division and then a BED is constructed such that the BED node is mapped to a cascade of quantum gates to construct the quantum circuit. Many decomposition methods \cite{2,16,sym13061025} are proposed to realize Toffoli gate with two controls in different forms based on the order of controls that decomposed the Toffoli gate with negative and positive controls into five elementary gates using eight different realizations.\\ 
\indent The aim of this paper is to synthesize the low-cost quantum circuit for a Boolean function represented as $PPRM$ expansion with $n$ variables. To achieve this aim, two optimization algorithms are proposed. The first algorithm generates an algebraic form to simplify the Boolean function and thus synthesis a low-cost quantum circuit. The second algorithm maps $MCT$ circuit into $NCV$ circuit using decomposition methods followed by applying a set of simplification rules. The proposed algorithms reduce the number of $MCT$ gates and thus reducing the quantum cost of quantum circuit.\\ 
\indent This paper is structured as follows. Sect.~\ref{sec:2} reviews the basic concepts of Boolean function and synthesis of reversible circuits using decomposition methods. Sect.~\ref{sec:prev method} reviews relevant methods for synthesizing quantum circuits. Sect.~\ref{sec:proposed method} presents the proposed algorithms to optimize the quantum circuits. Sect.~\ref{sec:result} presents the experimental results. Sect.~\ref{sec:Discussion} discusses the ways where the proposed algorithm can be used to build a stable medium scale and a small scale quantum computers and shows the synthesis of quantum circuits on IBM quantum computer. Sect.~\ref{sec:con} concludes the paper.
\section{Background}\label{sec:2}
\subsection{Boolean Functions} 
The Boolean function is a function that takes $n$ Boolean inputs and generates a unique Boolean output \cite{journals/qip/AhmedYE18}. For a Boolean function, each input vector maps to a single output vector \cite{journals/qip/AhmedYE18}. The Boolean function can be represented by Positive Polarity Reed-Muller ($PPRM$) as follows \cite{14}:

\begin{equation}
f(x_{1},..,x_{n})= \bigoplus_{i=1}^{2^{n}} a_{i} \beta_{i}, \beta_{i}= \Pi_{k=1}^{n} \quad x_{k}^{i_{k}},
\end{equation} 
where $a_{i}$ and $x_{k} \in \left\lbrace  0,1 \right\rbrace$ such that $a_{i}$ determines whether the product term exists or not. The multiplication indicates the AND operation, $\oplus$ is the XOR operation, $\beta_{i}$ is a product term and $i_{k}$ is the binary representation of $k$.  

\subsection{Boolean Quantum Circuits} 
\begin{figure}[!ht]
\centering
\begin{tabular}{ccc}
\Qcircuit @C=0.7em @R=0.5em @!R{
\lstick{}&	&\ctrlo{3}		&\qw		&\rstick{}\\
\lstick{}&	&\ctrl{2}		&\qw		&\rstick{}\\
\lstick{}&	&\qw			&\qw		&\rstick{}\\
\lstick{}&	&\targ		&\qw		&\rstick{}
}
&
\Qcircuit @C=0.7em @R=0.5em @!R{
\lstick{}&	&\ctrl{3}		&\qw		&\rstick{}\\
\lstick{}&	&\qw			&\qw		&\rstick{}\\
\lstick{}&	&\ctrl{1}		&\qw		&\rstick{}\\
\lstick{}&	&\targ		&\qw		&\rstick{}
}
&
\Qcircuit @C=0.7em @R=0.5em @!R{
\lstick{}&	&\ctrlo{3}		&\qw		&\rstick{}\\
\lstick{}&	&\ctrl{2}		&\qw		&\rstick{}\\
\lstick{}&	&\ctrl{1}		&\qw		&\rstick{}\\
\lstick{}&	&\targ		&\qw		&\rstick{}
}
\end{tabular}
\caption{$MCT$ gates with positive condition (closed circle) and negative condition (open circle)}
\label{fig:MCT gates}
\end{figure}
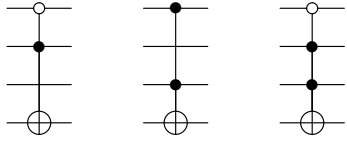

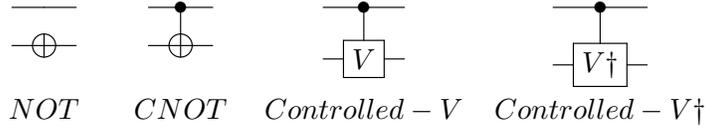
\begin{figure}[!ht]
\centering
\begin{tabular}{cccc}
\Qcircuit @C=0.7em @R=0.5em @!R{
\lstick{}&	&\qw			&\qw		&\rstick{}\\
\lstick{}&	&\targ		&\qw		&\rstick{}
}
&
\Qcircuit @C=0.7em @R=0.5em @!R{
\lstick{}&	&\ctrl{1}		&\qw		&\rstick{}\\
\lstick{}&	&\targ		&\qw		&\rstick{}
}
&
\Qcircuit @C=0.7em @R=0.5em @!R{
\lstick{}&	&\ctrl{1}			&\qw		&\rstick{}\\
\lstick{}&	&\gate{V}			&\qw		&\rstick{}
}
&
\Qcircuit @C=0.7em @R=0.5em @!R{
\lstick{}&	&\ctrl{1}			&\qw		&\rstick{}\\
\lstick{}&	&\gate{V\dagger}			&\qw		&\rstick{}
}
\\ [1cm] $NOT$  &$CNOT$ & $Controlled- V$ &  $Controlled-V\dagger$
\end{tabular}
\caption{$NCV$ quantum gate library}
\label{fig:NCV library}
\end{figure}

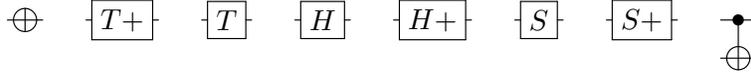
\begin{figure}[!ht]
\centering
\begin{tabular}{cccccccc}
\Qcircuit @C=0.2em @R=0.5em @!R{
\lstick{}&	&\targ		&\qw		&\rstick{}
}
&
\Qcircuit @C=0.2em @R=0.5em @!R{
\lstick{}&	&\gate{T+}		&\qw		&\rstick{}
}
&
\Qcircuit @C=0.2em @R=0.5em @!R{
\lstick{}&	&\gate{T}		&\qw		&\rstick{}
}
&
\Qcircuit @C=0.2em @R=0.5em @!R{
\lstick{}&	&\gate{H}		&\qw		&\rstick{}
}
&
\Qcircuit @C=0.2em @R=0.5em @!R{
\lstick{}&	&\gate{H+}		&\qw		&\rstick{}
}
&
\Qcircuit @C=0.2em @R=0.5em @!R{
\lstick{}&	&\gate{S}		&\qw		&\rstick{}
}
&
\Qcircuit @C=0.2em @R=0.5em @!R{
\lstick{}&	&\gate{S+}		&\qw		&\rstick{}
}
&
\Qcircuit @C=0.2em @R=0.5em @!R{
\lstick{}&	&\ctrl{1}		&\qw		&\rstick{}\\
\lstick{}&	&\targ		&\qw		&\rstick{}
}
\end{tabular}
\caption{$Clifford +T$ quantum gate library}
\label{fig:clifford library}
\end{figure}

The Boolean quantum circuit is a quantum circuit that realizes a Boolean function \cite{journals/qip/AhmedYE18}. 
Quantum circuits are composed of cascades of quantum gates \cite{2}. There are many quantum gate libraries to construct quantum circuits. The $MCT$, $NCV$ and $Clifford +T$ libraries are main quantum gate libraries that are adopted in this paper. The $MCT$ quantum gate library includes $CNOT$ gate with any number of controls. The $CNOT(C; t)$ denotes the $CNOT$ gate where $C$ is a set of control qubits and $t$ is a target qubit such that $t \notin C$. The number of control qubits is denoted by $\vert C \vert$. There are two conditions on control qubits, the positive condition and the negative condition. If control qubits have positive conditions, then the state of target qubit will be inverted if the values of control qubits are set to $ \vert 1 \rangle$ \cite{journals/qip/AhmedYE18}. If control qubits have negative conditions, then the state of target qubit will be inverted if the values of control qubits are set to $\vert 0 \rangle$ \cite{journals/qip/AhmedYE18}. Some $MCT$ gates with positive condition and negative condition are shown in Fig.~\ref{fig:MCT gates}. The $NOT$ gate is denoted by $CNOT(t)$. The $NCV$ quantum gate library in \cite{11} includes $CNOT(t)$, $CNOT(C; t)$ where $\vert C \vert=1$, $Controlled-V$ and $Controlled-V\dagger$ as shown in Fig.~\ref{fig:NCV library}. The $Clifford +T$ quantum gate library \cite{24} is composed of the quantum gates $T$, $T \dagger$, $S$, $S\dagger$, $NOT$, $H$, $H \dagger$ and $CNOT$ which is shown in Fig.~\ref{fig:clifford library}. Quantum circuit is called $MCT$ circuit if it is composed of gates from the $MCT$ library\cite{9,22}. $NCV$ circuit is a quantum circuit that is composed of gates from the $NCV$ library \cite{inbook}. $Clifford +T$  circuit is a quantum circuit that is composed of gates from the $Clifford +T$  library \cite{24}.

\subsection{ Decomposition of Quantum Circuits} 
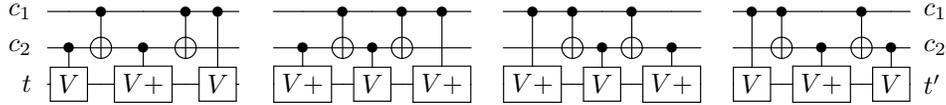
\begin{figure}[!ht]
\centering
\small\addtolength{\tabcolsep}{-3pt}
\begin{tabular}{ccccccccc}
\Qcircuit @C=0.1em @R=0.01em @!R{
\lstick{c_{1}}&	&\qw				&\ctrl{1}		&\qw				&\ctrl{1}		&\ctrl{1}			&\qw		&\rstick{}\\
\lstick{c_{2}}&	&\ctrl{1}			&\targ		&\ctrl{1}			&\targ		&\qw\qwx[1]		&\qw		&\rstick{}\\
\lstick{t}&	&\gate{V}			&\qw			&\gate{V+}		&\qw			&\gate{V}			&\qw		&\rstick{}
}
& &
\Qcircuit @C=0.01em @R=0.01em @!R{
\lstick{}&	&\qw				&\ctrl{1}		&\qw				&\ctrl{1}		&\ctrl{1}			&\qw		&\rstick{}\\
\lstick{}&	&\ctrl{1}			&\targ		&\ctrl{1}			&\targ		&\qw\qwx[1]		&\qw		&\rstick{}\\
\lstick{}&	&\gate{V+}		&\qw			&\gate{V}			&\qw			&\gate{V+}		&\qw		&\rstick{}
}
& &
\Qcircuit @C=0.01em @R=0.01em @!R{
\lstick{}&	&\ctrl{1}			&\ctrl{1}		&\qw				&\ctrl{1}		&\qw				&\qw		&\rstick{}\\
\lstick{}&	&\qw\qwx[1]		&\targ		&\ctrl{1}			&\targ		&\ctrl{1}			&\qw		&\rstick{}\\
\lstick{}&	&\gate{V+}		&\qw			&\gate{V}			&\qw			&\gate{V+}		&\qw		&\rstick{}
}
& &
\Qcircuit @C=0.01em @R=0.01em @!R{
\lstick{}&	&\ctrl{1}			&\ctrl{1}		&\qw				&\ctrl{1}		&\qw				&\qw		&\rstick{c_{1}}\\
\lstick{}&	&\qw\qwx[1]		&\targ		&\ctrl{1}			&\targ		&\ctrl{1}			&\qw		&\rstick{c_{2}}\\
\lstick{}&	&\gate{V}			&\qw			&\gate{V+}		&\qw			&\gate{V}			&\qw		&\rstick{t'}
}
\end{tabular}
\caption{Decomposition of $CNOT(c_{1},c_{2},t)$ with positive conditions on control qubits.}
\label{Decomposition positive CNOT}
\end{figure}

\begin{figure}[!ht]
\centering
\small\addtolength{\tabcolsep}{-3pt}
\begin{tabular}{ccccccccc}
\Qcircuit @C=0.01em @R=0.01em @!R{
\lstick{\bar{c}_{1}}&	&\qw				&\ctrlo{1}		&\qw				&\ctrlo{1}		&\ctrlo{1}			&\qw		&\rstick{}\\
\lstick{\bar{c}_{2}}&	&\ctrlo{1}			&\targ		&\ctrlo{1}			&\targ		&\qw\qwx[1]		&\qw		&\rstick{}\\
\lstick{t}&	&\gate{V}			&\qw			&\gate{V+}		&\qw			&\gate{V}			&\qw		&\rstick{}
}
& &
\Qcircuit @C=0.01em @R=0.01em @!R{
\lstick{}&	&\qw				&\ctrlo{1}		&\qw				&\ctrlo{1}		&\ctrlo{1}			&\qw		&\rstick{}\\
\lstick{}&	&\ctrlo{1}			&\targ		&\ctrlo{1}			&\targ		&\qw\qwx[1]		&\qw		&\rstick{}\\
\lstick{}&	&\gate{V+}		&\qw			&\gate{V}			&\qw			&\gate{V+}		&\qw		&\rstick{}
}
& &
\Qcircuit @C=0.01em @R=0.01em @!R{
\lstick{}&	&\ctrlo{1}			&\ctrlo{1}		&\qw				&\ctrlo{1}		&\qw				&\qw		&\rstick{}\\
\lstick{}&	&\qw\qwx[1]		&\targ		&\ctrlo{1}			&\targ		&\ctrlo{1}			&\qw		&\rstick{}\\
\lstick{}&	&\gate{V+}		&\qw			&\gate{V}			&\qw			&\gate{V+}		&\qw		&\rstick{}
}
& &
\Qcircuit @C=0.01em @R=0.01em @!R{
\lstick{}&	&\ctrlo{1}			&\ctrlo{1}		&\qw				&\ctrlo{1}		&\qw				&\qw		&\rstick{\bar{c}_{1}}\\
\lstick{}&	&\qw\qwx[1]		&\targ		&\ctrlo{1}			&\targ		&\ctrlo{1}			&\qw		&\rstick{\bar{c}_{2}}\\
\lstick{}&	&\gate{V}			&\qw			&\gate{V+}		&\qw			&\gate{V}			&\qw		&\rstick{t'}
}
\end{tabular}
\caption{Decomposition of $CNOT(\bar{c}_{1},\bar{c}_{2},t)$ with negative conditions on control qubits}
\label{Decomposition negative CNOT}
\end{figure}
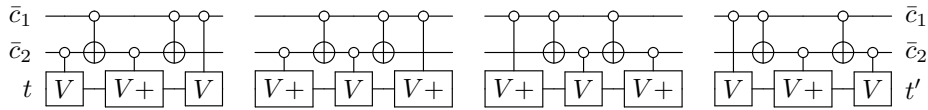

Decomposition of $MCT$ gate into its elementary gates has an important role to get an optimized quantum circuit by providing an opportunity to apply simplification rules which reduce number of quantum gates and thus the quantum cost of the quantum circuit. The $CNOT(C; t)$ gate with $\vert C \vert=2$ can be decomposed into five elementary gates \cite{16,9}. There are many realizations of decomposition of $CNOT$ gate with positive control qubits \cite{16} and negative control qubits \cite{sym13061025} as shown in Fig.~\ref{Decomposition positive CNOT} and Fig.~\ref{Decomposition negative CNOT} respectively. 

The basic definitions that are used later are reviewed as follows:
\begin{definition}
The quantum cost $QC$ of a quantum circuit is the total number of elementary quantum gates used to build the quantum circuit \cite{30,40}.
\end{definition}

\begin{definition}\label{def.homogenous}
For any homogeneous Boolean function with degree $d$, variables $n$ and product terms $p$ in the form, $\bigoplus_{l=1}^{p} x_{a}x_{b}..x_{h}x_{s_{l}}$,
where $ 1 \leq d \leq n$ and each product term is of degree $d$; it can be factorized to the form \cite{5}, $(x_{a}x_{b}..x_{h})\bigoplus_{l=1}^{p} x_{s_{l}}$, where $x_{a}x_{b}..x_{h}$ is the factor group and $(\bigoplus_{l=1}^{p}x_{s_{l}})$ is the corresponding factor variables. The quantum circuit of the Boolean function is a set of ordered $CNOT$ gates as follows:
\small
\begin{equation}\label{circuit fac}
\centering
\begin{aligned}
&CNOT(x_{s_{1}};x_{s_{2}})CNOT(x_{s_{2}};x_{s_{3}}) .. CNOT(x_{s_{l-1}};x_{s_{l}})CNOT(x_{a},x_{b},..,x_{h},x_{s_{l}};f) \\ & 
CNOT(x_{s_{l-1}};x_{s_{l}}) .. CNOT(x_{s_{2}};x_{s_{3}})CNOT(x_{s_{1}};x_{s_{2}}),
\end{aligned}
\end{equation}
\normalsize
where the result of the Boolean function will be stored on $x_{s_{l}}$. The numbers of variables in factor variables are denoted by $LenF$. It should be noted that the last $(LenF-1)$ $CNOT$ gates can be removed from a quantum circuit in Eq.\eqref{circuit fac} without change in result of quantum circuit where the input qubits will be assumed as garbage.
\end{definition}

\begin{definition}\label{CTR rule def}
For any two $CNOT_{1}(C_{1}; t)$ and $CNOT_{2}(C_{2}; t)$ gates with positive conditions on control qubits and the common target, if $C_{1} \subset C_{2}$ and $\vert C_{2}|-|C_{1}\vert =1$, then apply the Common-Target Rule $(CTR)$ \cite{Arabzadeh:2010:ROR:1899721.1899916} and merge these two gates into one $CNOT_{r}(C_{r}; t)$ gate where $C_{r}$ is the set of control qubits of $CNOT_{1}$ with positive conditions and different control qubit between $C_{1}$ and $C_{2}$ with negative condition as shown in Fig.~\ref{CTR fig}
\end{definition}
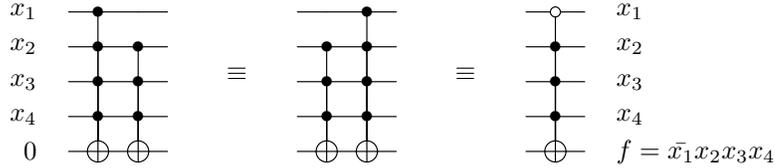
\begin{figure}[!ht]
\centering
\small\addtolength{\tabcolsep}{-10pt}
\begin{tabular}{ccccc}
\Qcircuit @C=0.7em @R=0.5em @!R{
\lstick{x_{1}}&	&\ctrl{4}		&\qw		&\qw		&\rstick{}\\
\lstick{x_{2}}&	&\ctrl{3}		&\ctrl{3}		&\qw		&\rstick{}\\
\lstick{x_{3}}&	&\ctrl{2}		&\ctrl{2}		&\qw		&\rstick{}\\
\lstick{x_{4}}&	&\ctrl{1}		&\ctrl{1}			&\qw		&\rstick{}\\
\lstick{0}&	&\targ		&\targ		&\qw		&\rstick{}
}
& \raisebox{-6ex}{ \, \, \, \, $ \equiv $  \, \, \, \,}&
\Qcircuit @C=0.7em @R=0.5em @!R{
\lstick{}&	&\qw		&\ctrl{4}		&\qw		&\rstick{}\\
\lstick{}&	&\ctrl{3}		&\ctrl{3}		&\qw		&\rstick{}\\
\lstick{}&	&\ctrl{2}		&\ctrl{2}		&\qw		&\rstick{}\\
\lstick{}&	&\ctrl{1}			&\ctrl{1}		&\qw		&\rstick{}\\
\lstick{}&	&\targ		&\targ		&\qw		&\rstick{}
}
& \raisebox{-6ex}{ \, \, \, \, $ \equiv $  \, \, \, \,}&
\Qcircuit @C=0.7em @R=0.5em @!R{
\lstick{}&	&\ctrlo{4}		&\qw		&\rstick{x_{1}}\\
\lstick{}&	&\ctrl{3}		&\qw		&\rstick{x_{2}}\\
\lstick{}&	&\ctrl{2}		&\qw		&\rstick{x_{3}}\\
\lstick{}&	&\ctrl{1}		&\qw		&\rstick{x_{4}}\\
\lstick{}&	&\targ		&\qw		&\rstick{f=\bar{x_{1}}x_{2}x_{3}x_{4}}
}
\end{tabular}
\caption{Common-Target Rule (CTR) \cite{Arabzadeh:2010:ROR:1899721.1899916}}
\label{CTR fig}
\end{figure}

\section{Background Methods} 
\label{sec:prev method}
Synthesizing a reversible Boolean circuit with minimal quantum cost depends on the form and the number of terms of the corresponding Boolean function. This section reviews three methods that we will be used by the proposed algorithm to synthesize a quantum circuit. The first method is the direct method \cite{14} that constructs a reversible circuit for any Boolean function. The factorization algorithm \cite{5} is the second method to construct a reversible circuit for any homogeneous Boolean function. The third method discussed in this section is called the reorder method \cite{journals/qip/AhmedYE18} that applies the factorization algorithm \cite{5} and then rewrites the terms of Boolean function.
\subsection{The Direct Method}
The reversible circuit of a Boolean function with $n$ variables of product terms can be constructed by the following steps \cite{14}:
\begin{enumerate}
\item Prepare $(n+1)$ qubits and set the last qubit to $\vert 0 \rangle$ to store the result of the function.
\item Add $CNOT(C; t)$ gate with control qubits of positive conditions for each product term of the Boolean function $f$ in the form $ x_{k_{1}}x_{k_{2}}..x_{k_{n}}$ where $C=\left\lbrace x_{k_{1}}, x_{k_{2}}, .. ,x_{k_{n}}\right\rbrace $ and the result can be stored on the target qubit $t$.
\item Add $CNOT(C; t)$ gate with the control qubits of negative conditions for each product term of the Boolean function $f$ in the form $ \bar{x}_{k_{1}}\bar{x}_{k_{2}}..\bar{x}_{k_{n}} $ where $C=\left\lbrace \bar{x}_{k_{1}}, \bar{x}_{k_{2}}, .. ,\bar{x}_{k_{n}}\right\rbrace $ and the result can be stored on the target qubit $t$.
\item Add $CNOT(t)$ gate for each product term is equal to 1.
\end{enumerate}
For example, a Boolean function with four variables is considered in Eq.\eqref{eq:example direct method}, and the synthesized quantum circuit using the direct method \cite{14} is shown in Fig.~\ref{f.direct method}.
\begin{equation} 
\label{eq:example direct method}
f(x_{1},x_{2},x_{3},x_{4})= x_{1}x_{2}x_{3}x_{4}\oplus \bar{x}_{1}x_{2}x_{4}\oplus \bar{x}_{2}x_{3}x_{4}\oplus x_{3}x_{4}\oplus 1.
\end{equation}
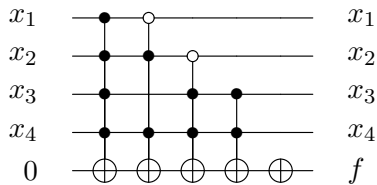
\begin{figure}[!ht]
\[
\Qcircuit @C=0.7em @R=0.5em @!R{
\lstick{x_{1}}&	&\ctrl{4}		&\ctrlo{4}		&\qw			&\qw			&\qw			&\qw		&\rstick{x_{1}}\\
\lstick{x_{2}}&	&\ctrl{3}		&\ctrl{3}		&\ctrlo{3}		&\qw			&\qw			&\qw		&\rstick{x_{2}}\\
\lstick{x_{3}}&	&\ctrl{2}		&\qw			&\ctrl{2}		&\ctrl{2}		&\qw			&\qw		&\rstick{x_{3}}\\
\lstick{x_{4}}&	&\ctrl{1}		&\ctrl{1}		&\ctrl{1}		&\ctrl{1}		&\qw			&\qw		&\rstick{x_{4}}\\
\lstick{0}&	&\targ		&\targ		&\targ		&\targ		&\targ		&\qw		&\rstick{f}
}
\]
\caption{The synthesized quantum circuit using the direct method \cite{14} of the Boolean function in Eq.\eqref{eq:example direct method}}
\label{f.direct method}
\end{figure}
\subsection{The Factorization Algorithm} 
According to \cite{5}, a factorization algorithm factorizes the product terms of the Boolean function $f$ and uses $p$ and $n$ to denote the product term and number of variables of the Boolean function respectively. $FactTable(G)$ is a factorization table with $n$ rows and $p$ columns for each group $G$ of product terms with the same degree $d$. $FactTable(G)$ can be filled row by row; the value of the cell will be set to 1 if the variable exists and will be set to 0 otherwise. The following steps can be applied until the factorization table is empty:
\begin{enumerate}
\item Sum all values for each variable (column) and choose the variable with the first maximum value as a factor variable $x_{s}$, then the Boolean function can be represented in the form, $(fg_{s_{1}} \oplus .. \oplus fg_{s_{n}})x_{s}\oplus R,$   
where $x_{s}$ is a factor variable and its factor group (product term) is $ fg_{s_{1}} \oplus .. \oplus fg_{s_{n}}$ and $R$ is the remainder terms of function. 
\item Remove all rows that include the factor variable $x_{s}$.
\item Collect all factor variables with their factor group by applying steps (1) and (2) on the factorization table.
\end{enumerate}  

$AnalysisCommonFactor(G)$ analyzes all factor groups to find common terms $fg$ to get the form in Definition \ref{def.homogenous} and collects common terms $fg$ for all factor variables $x_{s}$ to get the form of each group of product terms with the same degree,
     \begin{equation}
     \begin{aligned}
    & fg_{1}(x_{s_{0}}\oplus .. \oplus x_{s_{n}})\oplus fg_{2}(x_{s_{0}} \oplus .. \oplus x_{s_{n}}\oplus x_{s_{n+1}}) \\ &\oplus .. \oplus fg_{h}(x_{s_{0}}\oplus .. \oplus x_{s_{n}}        
      \oplus ..\oplus x_{s_{n+h+1}}).                 
     \end{aligned}
      \end{equation}
Then, we can get the function after factorization $f_{z}$ by applying $XOR$ operation among all groups as in Eq.\eqref{fac}. Finally, all terms of $f_{z}$ are either product terms or in the form [factor group][factor variables] which will be referred to as $[g][v]$. The quantum circuit is synthesized for all $[g][v]$ terms as in Definition~\ref{def.homogenous}. For all product terms, the quantum circuit is constructed using the direct method \cite{14}.
\begin{equation}
     \begin{aligned}
      & f_{z} = G_{1} \oplus G_{2} \oplus .. \oplus G_{n}.                
     \end{aligned}
     \label{fac}
\end{equation}

\subsection{The Reorder Method}
According to \cite{journals/qip/AhmedYE18}, the reorder method factorizes the reversible Boolean function using the factorization algorithm \cite{5} and then applies two stages on the reversible Boolean function such that all product terms can be sorted in ascending order based on degree $d$ to apply the $CTR$ if possible. The first stage presents all possible forms for the factor group $[g]$ and applies a specific order for the factor group $[g]$ for all $[g][v]$ terms that have the same $LenF$ as follows:

  $F_{1}$ : the XOR of product terms $x_{s_{1}}x_{s_{2}}..x_{s_{i}}\oplus..\oplus x_{s_{n}}x_{s_{m}}..x_{s_{i}}$
  
  $F_{2}$ : a product term $x_{s_{1}}x_{s_{2}}..x_{s_{i}}$

  $F_{3}$ : a positive literal $x_{s_{n}}$ 

  $F_{4}$ : the $[g][v]$ form such that its factor group $[g]$ is a positive literal \indent \indent $(x_{s_{i}})(x_{s_{1}}\oplus x_{s_{2}}\oplus..\oplus x_{s_{n}})$ 

  $F_{5}$ : the XOR of positive literals  $x_{s_{1}}\oplus x_{s_{2}}\oplus..\oplus x_{s_{n}}$\\
Then all $[g][v]$ terms can be sorted in ascending order based on $LenF$. The second stage applies a set of reorder rules to reorder and simplify the terms of the function as shown in Table.~\ref{tab:reorder Rules}. After applying reorder rules, all $[g][v]$ terms with common $[v]$ with different $LenF$ can be collected as a group. For each group, the order of forms $F_{1}, F_{2}, F_{3}, F_{4}$ and $F_{5}$ can be applied on all factor groups $[g]$ of $[g][v]$ terms with common $[v]$ with the same $LenF$. For each group, the maximum $LenF$ can be calculated and all $[g][v]$ terms can be sorted in ascending order based on $LenF$. The second stage is finished by sorting all groups in ascending order based on the maximum $LenF$. 
\begin{table}[htb]
\centering
\small\addtolength{\tabcolsep}{-1pt}
\begin{tabular}{ccc}
\hline 
Rule \# & Specification & Example \\ 
\hline \\[-1em]
$R_{1}$ & \begin{tabular}[c]{@{}c@{}}If the factor group for \\any term in form $[g][v]$ \\ is the same as another term\\ $T1$ in the Boolean function,\\ then merge two terms into \\ one term in the form \\ $(T1)$(factor variables $\oplus 1$) \end{tabular} & \begin{tabular}[c]{@{}c@{}}$x_{1}x_{2}\oplus (x_{1}x_{2})(x_{3}\oplus x_{4})$\\ $=(x_{1}x_{2})(x_{3}\oplus x_{4}\oplus 1)$\\or\\$(x_{1}(x_{2} \oplus x_{3}))(x_{6} \oplus x_{7}) \oplus x_{1}(x_{2} \oplus x_{3})$\\ $=(x_{1}(x_{2} \oplus x_{3}))(x_{6} \oplus x_{7}\oplus 1)$.\end{tabular} \\ \\ 
$R_{2}$ & \begin{tabular}[c]{@{}c@{}}If number of variables in \\ factor group $[g]$ that is in \\$F_{4}$ or $F_{5}$ is greater than \\$LenF,$ then the factor group \\and the factor variables can\\ be exchanged \end{tabular} & \begin{tabular}[c]{@{}c@{}}$(x_{1}(x_{2} \oplus x_{3} \oplus x_{4} \oplus x_{5}))(x_{6} \oplus x_{7})$ \\ $=(x_{1}(x_{6} \oplus x_{7}))(x_{2} \oplus x_{3} \oplus x_{4} \oplus x_{5})$. \end{tabular} \\  
\hline
\end{tabular}
\caption{The reorder rules of the reorder algorithm \cite{journals/qip/AhmedYE18}.}
\label{tab:reorder Rules}
\end{table}

\section{The Proposed Algorithm}
\label{sec:proposed method}
The aim of the proposed algorithm is to minimize the quantum cost of a quantum circuit that realizes a given Boolean function represented as a Positive Polarity Reed-Muller $PPRM$ expansion with $n$ variables. The given Boolean function has various algebraic forms and each algebraic form of them can be realized by a quantum circuit with a different quantum cost. So, Synthesis of quantum circuit with minimal quantum cost for a given Boolean function depends on the algebraic form of the Boolean function.
The proposed algorithm has two main steps to construct a quantum circuit that achieves minimizing the quantum cost. The first step is generating an algebraic form for a given Boolean function represented as $PPRM$ expansion then constructing the $MCT$ circuit by synthesis methods discussed in Sect.~\ref{sec:prev method}. The second step is mapping the $MCT$ circuit into the $NCV$ circuit.
\subsection{Formation of an Algebraic Form}
There are various algebraic forms that can be generated for any Boolean function, and therefore there are various quantum circuits that can be synthesized for a Boolean function with different quantum cost depends on its algebraic form. So, the first step to construct a quantum circuit to minimize quantum cost is generating a simple algebraic form for the Boolean function. This paper proposes a special algebraic form by rearrangement of terms of a given Boolean function expressed in $PPRM$ based on degree of term $d_{term}$ that can be calculated as shown in Table.~\ref{degree term table}. Firstly, the proposed algorithm applies the factorization algorithm \cite{5} on the Boolean function and then applies the reorder method \cite{journals/qip/AhmedYE18} on the factorized Boolean function. Secondly, the terms of the Boolean function can rewritten by the proposed special algebraic form to get an algebraic form for the Boolean function that can realized by the quantum circuit with minimal quantum cost. Finally, the quantum circuit of the Boolean function can be synthesized using the synthesis methods. 
\begin{table}[!ht]
\centering
\small\addtolength{\tabcolsep}{-1pt}
\begin{tabular}{cccc}
\hline \\[-1em] 
 Case\# & Description of the term & $d_{term}$      \\
\hline \\[-1ex] 
 Case 1 & is a product term $x_{1}x_{2} .. x_{n}$ & $d_{term} = n$  
 \\ [3ex]
 Case 2 & \begin{tabular}[c]{@{}c@{}} is in $[g][v]$ form and the factor \\ group is in $F_{1}$ as $p_{1}\oplus p_{2} \oplus p_{i}$ \end{tabular} & \begin{tabular}[c]{@{}c@{}} $d_{term}$  = $max(d_{term}$ \\for each product term $p)$ \end{tabular}
 \\ [3ex]
 Case 3 & \begin{tabular}[c]{@{}c@{}} is in $[g][v]$ form  and the factor \\group is in $F_{2}$ as $x_{1}x_{2} .. x_{n}$ \end{tabular} & $d_{term} = n + 1$ 
 \\ [3ex]
 Case 4 & \begin{tabular}[c]{@{}c@{}} is in $[g][v]$ form  and the factor \\group is in $F_{3}$ as $x_{i}$ \end{tabular} & $d_{term} = 2$ 
 \\ [3ex]
 Case 5 & \begin{tabular}[c]{@{}c@{}} is in $[g][v]$ form and the factor \\ group is in $F_{4}$ as \\$x_{i}(x_{1}\oplus x_{2}\oplus ..\oplus x_{n})$ \end{tabular} & $d_{term} = 3$ 
 \\ [4ex]
  Case 6 & \begin{tabular}[c]{@{}c@{}} is in $[g][v]$ form and the factor \\group is in $F_{5}$ as $x_{1} \oplus x_{2}\oplus ..\oplus x_{n}$ \end{tabular} & $d_{term} = 2$ 
\\ \hline                
\end{tabular}
\caption{The degree of term for all possible cases of terms of Boolean function.}
\label{degree term table}
\end{table}

\begin{figure}[!ht]
\centering
\begin{tabular}{ccccc}
\Qcircuit @C=0.3em @R=0.5em @!R{
\lstick{x_{1}}&	&\ctrl{5}		&\qw			&\ctrl{5}		&\qw			&\qw		&\rstick{x_{1}}\\
\lstick{x_{2}}&	&\ctrl{4}		&\qw			&\qw			&\qw			&\qw		&\rstick{x_{2}}\\
\lstick{x_{3}}&	&\ctrl{3}		&\ctrl{2}		&\qw			&\ctrl{2}		&\qw		&\rstick{x_{3}}\\
\lstick{x_{4}}&	&\ctrl{2}		&\qw			&\qw			&\qw			&\qw		&\rstick{x_{4}}\\
\lstick{x_{5}}&	&\qw			&\targ		&\ctrl{1}		&\targ		&\qw		&\rstick{x_{5}}\\
\lstick{0}&	&\targ		&\qw			&\targ		&\qw			&\qw		&\rstick{f}
}
&  \raisebox{-8ex}{\,\,\,\,\,\ $\Rightarrow $ }   &
\Qcircuit @C=0.3em @R=0.5em @!R{
\lstick{}&	&\qw			&\ctrl{4}		&\qw			&\ctrl{4}		&\qw			&\ctrl{5}		&\qw			&\qw		&\rstick{x_{1}}\\
\lstick{}&	&\qw			&\ctrl{3}		&\qw			&\ctrl{3}		&\qw			&\qw			&\qw			&\qw		&\rstick{x_{2}}\\
\lstick{}&	&\qw			&\ctrl{2}		&\qw			&\ctrl{2}		&\ctrl{2}		&\qw			&\ctrl{2}		&\qw		&\rstick{x_{3}}\\
\lstick{}&	&\ctrl{2}		&\qw			&\ctrl{2}		&\qw			&\qw			&\qw			&\qw			&\qw		&\rstick{x_{4}}\\
\lstick{}&	&\ctrl{1}		&\targ		&\ctrl{1}		&\targ		&\targ		&\ctrl{1}		&\targ		&\qw		&\rstick{x_{5}}\\
\lstick{}&	&\targ		&\qw			&\targ		&\qw			&\qw			&\targ		&\qw			&\qw		&\rstick{f}
}
&  \raisebox{-8ex}{\,\,\,\,\,\ $\Rightarrow $ }   &
\Qcircuit @C=0.3em @R=0.5em @!R{
\lstick{}&	&\qw			&\ctrl{4}		&\qw			&\ctrl{4}		&\qw			&\ctrl{5}		&\qw		&\rstick{gr}\\
\lstick{}&	&\qw			&\ctrl{3}		&\qw			&\ctrl{3}		&\qw			&\qw			&\qw		&\rstick{gr}\\
\lstick{}&	&\qw			&\ctrl{2}		&\qw			&\ctrl{2}		&\ctrl{2}		&\qw			&\qw		&\rstick{gr}\\
\lstick{}&	&\ctrl{2}		&\qw			&\ctrl{2}		&\qw			&\qw			&\qw			&\qw		&\rstick{gr}\\
\lstick{}&	&\ctrl{1}		&\targ		&\ctrl{1}		&\targ		&\targ		&\ctrl{1}		&\qw		&\rstick{gr}\\
\lstick{}&	&\targ		&\qw			&\targ		&\qw			&\qw			&\targ		&\qw		&\rstick{f}
}
\\
\\
(a) && (b) && (c)
\end{tabular}
\caption{Reversible circuit of $f_{1}$ where (a) represents $f_{1}$ in Eq.\eqref{f1}, (b) represents $f_{1}$ after decomposing the first gate without remove any gate and (c) represents $f_{1}$ after decomposing the first gate and remove the last gate that has no effect on the result of $f_{1}$ where $g$ is a garbage qubit with $QC=42$.}
\label{circuit f1 before decom}
\end{figure}
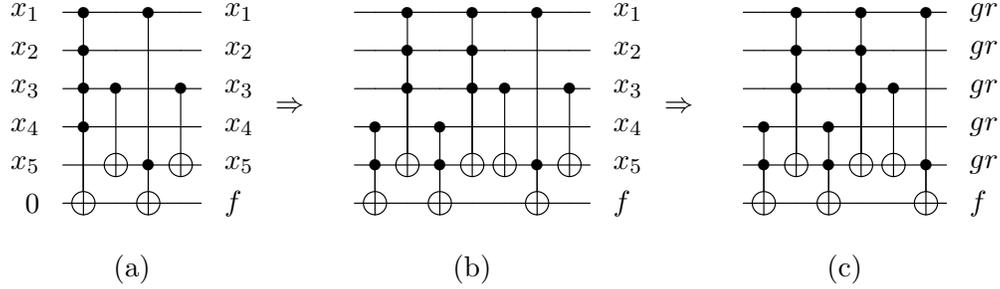

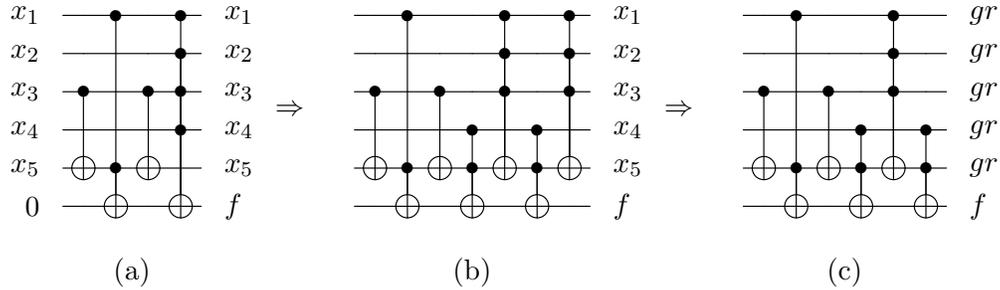
\begin{figure}[!ht]
\centering
\begin{tabular}{ccccc}
\Qcircuit @C=0.3em @R=0.5em @!R{
\lstick{x_{1}}&	&\qw			&\ctrl{5}		&\qw			&\ctrl{5}		&\qw		&\rstick{x_{1}}\\
\lstick{x_{2}}&	&\qw			&\qw			&\qw			&\ctrl{4}		&\qw		&\rstick{x_{2}}\\
\lstick{x_{3}}&	&\ctrl{2}		&\qw			&\ctrl{2}		&\ctrl{3}		&\qw		&\rstick{x_{3}}\\
\lstick{x_{4}}&	&\qw			&\qw			&\qw			&\ctrl{2}		&\qw		&\rstick{x_{4}}\\
\lstick{x_{5}}&	&\targ		&\ctrl{1}		&\targ		&\qw			&\qw		&\rstick{x_{5}}\\
\lstick{0}&	&\qw			&\targ		&\qw			&\targ		&\qw		&\rstick{f}
}
&  \raisebox{-8ex}{\,\,\,\,\,\ $\Rightarrow $ }   &
\Qcircuit @C=0.3em @R=0.5em @!R{
\lstick{}&	&\qw			&\ctrl{5}		&\qw			&\qw			&\ctrl{4}		&\qw			&\ctrl{4}		&\qw		&\rstick{x_{1}}\\
\lstick{}&	&\qw			&\qw			&\qw			&\qw			&\ctrl{3}		&\qw			&\ctrl{3}		&\qw		&\rstick{x_{2}}\\
\lstick{}&	&\ctrl{2}		&\qw			&\ctrl{2}		&\qw			&\ctrl{2}		&\qw			&\ctrl{2}		&\qw		&\rstick{x_{3}}\\
\lstick{}&	&\qw			&\qw			&\qw			&\ctrl{2}		&\qw			&\ctrl{2}		&\qw			&\qw		&\rstick{x_{4}}\\
\lstick{}&	&\targ		&\ctrl{1}		&\targ		&\ctrl{1}		&\targ		&\ctrl{1}		&\targ		&\qw		&\rstick{x_{5}}\\
\lstick{}&	&\qw			&\targ		&\qw			&\targ		&\qw			&\targ		&\qw			&\qw		&\rstick{f}
}
&  \raisebox{-8ex}{\,\,\,\,\,\ $\Rightarrow $ }   &
\Qcircuit @C=0.3em @R=0.5em @!R{
\lstick{}&	&\qw			&\ctrl{5}		&\qw			&\qw			&\ctrl{4}		&\qw			&\qw		&\rstick{gr}\\
\lstick{}&	&\qw			&\qw			&\qw			&\qw			&\ctrl{3}		&\qw			&\qw		&\rstick{gr}\\
\lstick{}&	&\ctrl{2}		&\qw			&\ctrl{2}		&\qw			&\ctrl{2}		&\qw			&\qw		&\rstick{gr}\\
\lstick{}&	&\qw			&\qw			&\qw			&\ctrl{2}		&\qw			&\ctrl{2}		&\qw		&\rstick{gr}\\
\lstick{}&	&\targ		&\ctrl{1}		&\targ		&\ctrl{1}		&\targ		&\ctrl{1}		&\qw		&\rstick{gr}\\
\lstick{}&	&\qw			&\targ		&\qw			&\targ		&\qw			&\targ		&\qw		&\rstick{f}
}
\\
\\
(a) && (b) && (c)
\end{tabular}
\caption{Reversible circuit of $f_{2}$ where (a) represents $f_{2}$ in Eq.\eqref{f2}, (b) represents $f_{2}$ after decomposing the last gate without removing any gate and (c) represents $f_{2}$ after decomposing the last gate and removing the last gate that has no effect on the result of $f_{2}$ where $g$ is a garbage qubit with $QC=30$.}
\label{circuit f2 before decom}
\end{figure}

To get an algebraic form for the Boolean function by proposed rearrangement of terms, the term of the Boolean function that has the maximum $d_{term}$ will be the last term in the Boolean function. For example, consider a Boolean function $f_{1}$ after applying the reorder method \cite{journals/qip/AhmedYE18} in Eq.\eqref{f1} with two terms, $x_{1}x_{2}x_{3}x_{4}$ and $x_{1}(x_{3} \oplus x_{5})$, and their $d_{term}$ are 4 and 2 respectively. The Boolean function $f_{2}$ in Eq.\eqref{f2} is the same as $f_{1}$ after applying the proposed rearrangement of terms of $f_{1}$ based on values of $d_{term}$ of terms such that the term with the maximum $d_{term}$ will be the last term in the Boolean function $f_{2}$. The quantum circuits of Boolean functions $f_{1}$ and $f_{2}$ are shown in Fig.~\ref{circuit f1 before decom} and Fig.~\ref{circuit f2 before decom} respectively. The quantum cost $QC$ of quantum circuits that realize $f_{1}$ and $f_{2}$ after the decomposition of the $CNOT(x_{1},x_{2},x_{3},x_{4},f)$ gate is 43. After ignoring the last gate that restores the state of the inputs, the $QC$ of $f_{1}$ is reduced to 42 and $QC$ of $f_{2}$ is reduced to 30.
\begin{equation}
f_{1}= x_{1}x_{2}x_{3}x_{4} \oplus x_{1}(x_{3} \oplus x_{5})
\label{f1}
\end{equation}

\begin{equation}
f_{2}= x_{1}(x_{3} \oplus x_{5}) \oplus x_{1}x_{2}x_{3}x_{4}
\label{f2}
\end{equation}

\begin{algorithm}[!ht]
\caption{Formation an Algebraic Form}\label{strategy1}
\DontPrintSemicolon
\SetAlgoNoLine%
       $f \gets $ Factorization Algorithm($f$)\\
       $f\gets$ Reorder Method($f)$\\
       \For {term in $f$}
       {
       \uIf{term is Case 1}{$d_{term} \gets n$}
       \uElseIf{term is Case 2}{$d_{term} \gets$ max($d_{term}$ for each product term)}
       \uElseIf{term is Case 3}{$d_{term} \gets n+1$}
       \uElseIf{term is Case 4 or Case 6}{$d_{term} \gets 2$}
       \ElseIf{term is Case 5}{$d_{term} \gets 3$} 
       }
       $T \gets$ term of maximum $d_{term}$\\
       Remove $T$ from $f$\\
       $f_{new} \gets f \oplus T$\\
       \For {product term in $f_{new}$}
       {Apply the Direct Method}
       \For{$[g][v]$ term in $f_{new}$}
       {Synthesize the quantum circuit as in Definition~\ref{def.homogenous} }
\end{algorithm}

\subsection{Mapping MCT Circuit into NCV Circuit}
\subsubsection{Basic MCT Gates Decomposition}
According to the basic decomposition method \cite{9}, the number of qubits of a quantum circuit is denoted by $w$. There are three basic decomposition rules of $CNOT(C; t)$ gate with $\vert C \vert$ of control qubits that called DR1, DR2 and DR3 as shown in Table.~\ref{decomposition rules}. The first decomposition rule (DR1) decomposes a $CNOT(C; t)$ into $4(\vert C \vert-2)$ $CNOT(C; t)$ gates, if $\vert C \vert \in \lbrace 3,..,\lceil w/2 \rceil\rbrace$ and $w\geq5$. If $w \geq 5$ and $\vert C \vert >\lceil w/2 \rceil$, the second decomposition rule (DR2) is applied to decompose a $CNOT(C; t)$ gate into four $CNOT$ gates with $\lceil w/2 \rceil$ and $w-1-\lceil w/2\rceil$ control qubits. The third decomposition rule (DR3) is applied on a $CNOT(C; t)$ gate when $\vert C \vert > 2$ and $\vert C \vert = w-1$ by adding the auxiliary line to $CNOT$ gate and then applying the second decomposition rule (DR2).
\subsubsection{Optimization by Simplification Rules}
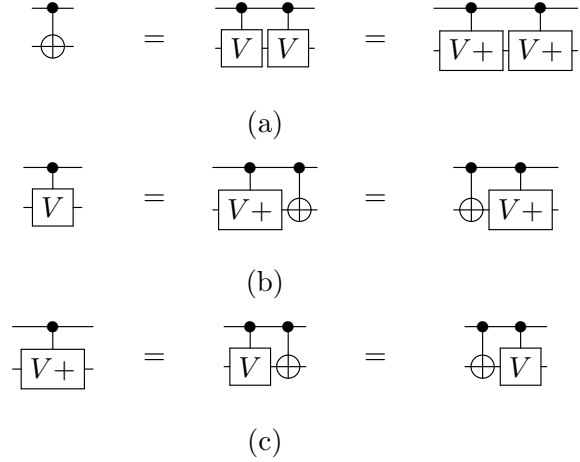
\begin{figure}[!ht]
\centering
\begin{tabular}{ccccc}
\Qcircuit @C=0.3em @R=0.5em @!R{
\lstick{}&	&\ctrl{1}		&\qw		&\rstick{}\\
\lstick{}&	&\targ		&\qw		&\rstick{}
}
&  \raisebox{-3ex}{ $=$ }   &
\Qcircuit @C=0.2em @R=0.1em @!R{
\lstick{}&	&\ctrl{1}			&\ctrl{1}			&\qw		&\rstick{}\\
\lstick{}&	&\gate{V}			&\gate{V}			&\qw		&\rstick{}
}
&  \raisebox{-3ex}{ $=$ }   &
\Qcircuit @C=0.2em @R=0.1em @!R{
\lstick{}&	&\ctrl{1}			&\ctrl{1}			&\qw		&\rstick{}\\
\lstick{}&	&\gate{V+}			&\gate{V+}			&\qw		&\rstick{}
}
\\ \\
& & (a)&  &
\\
\Qcircuit @C=0.3em @R=0.1em @!R{
\lstick{}&	&\ctrl{1}			&\qw		&\rstick{}\\
\lstick{}&	&\gate{V}			&\qw		&\rstick{}
}
&  \raisebox{-3ex}{ $=$ }   &
\Qcircuit @C=0.2em @R=0.1em @!R{
\lstick{}&	&\ctrl{1}			&\ctrl{1}		&\qw		&\rstick{}\\
\lstick{}&	&\gate{V+}		&\targ		&\qw		&\rstick{}
}
&  \raisebox{-3ex}{ $=$ }   &
\Qcircuit @C=0.2em @R=0.1em @!R{
\lstick{}&	&\ctrl{1}		&\ctrl{1}			&\qw		&\rstick{}\\
\lstick{}&	&\targ		&\gate{V+}		&\qw		&\rstick{}
}
\\ \\
& & (b)&  &
\\
\Qcircuit @C=0.3em @R=0.1em @!R{
\lstick{}&	&\ctrl{1}			&\qw		&\rstick{}\\
\lstick{}&	&\gate{V+}			&\qw		&\rstick{}
}
&  \raisebox{-3ex}{ $=$ }   &
\Qcircuit @C=0.2em @R=0.1em @!R{
\lstick{}&	&\ctrl{1}			&\ctrl{1}		&\qw		&\rstick{}\\
\lstick{}&	&\gate{V}		&\targ		&\qw		&\rstick{}
}
&  \raisebox{-3ex}{ $=$ }   &
\Qcircuit @C=0.2em @R=0.1em @!R{
\lstick{}&	&\ctrl{1}		&\ctrl{1}			&\qw		&\rstick{}\\
\lstick{}&	&\targ		&\gate{V}		&\qw		&\rstick{}
}
\\ \\
& & (c)&  &
\end{tabular}
\caption{Merge Rules}
\label{optimization rules fig}
\end{figure}

The $NCV$ circuit can be simplified by reducing the number of gates using the following simplification rules:

\begin{Rule}
If two  $Controlled-V$ gates or  $Controlled-V\dagger$ gates are adjacent, then these two gates can be merged into $CNOT$ gate as shown in Fig.~\ref{optimization rules fig}(a).
\end{Rule}

\begin{table}[H]
\centering
\small\addtolength{\tabcolsep}{-3pt}
\begin{tabular}{ccc}
\hline
 Rule\# & Description     & Examples \\
\hline \\[-1em]                              
DR1   & \begin{tabular}[c]{@{}c@{}}if $w\geq5$ and $\vert C \vert \in \lbrace 3,..,\lceil w/2 \rceil\rbrace$,\\ then $CNOT(C; t)$ is decomposed \\ into $4(\vert C \vert -2)$ $CNOT(C; t)$ gates \end{tabular}& 
\begin{tabular}{ccc}
\Qcircuit @C=0.01em @R=0.5em @!R{
\lstick{}&	&\ctrl{8}		&\qw		&\rstick{}\\
\lstick{}&	&\ctrl{7}		&\qw		&\rstick{}\\
\lstick{}&	&\ctrl{6}		&\qw		&\rstick{}\\
\lstick{}&	&\ctrl{5}		&\qw		&\rstick{}\\
\lstick{}&	&\ctrl{4}		&\qw		&\rstick{}\\
\lstick{}&	&\qw			&\qw		&\rstick{}\\
\lstick{}&	&\qw			&\qw		&\rstick{}\\
\lstick{}&	&\qw			&\qw		&\rstick{}\\
\lstick{}&	&\targ		&\qw		&\rstick{}
}
&  \raisebox{-15ex}{  $ \equiv $ }   &
\Qcircuit @C=0.01em @R=0.5em @!R{
\lstick{}&	&\qw			&\qw			&\qw			&\ctrl{5}		&\qw			&\qw			&\qw			&\qw			&\qw			&\ctrl{5}		&\qw			&\qw			&\qw		&\rstick{}\\
\lstick{}&	&\qw			&\qw			&\qw			&\ctrl{4}		&\qw			&\qw			&\qw			&\qw			&\qw			&\ctrl{4}		&\qw			&\qw			&\qw		&\rstick{}\\
\lstick{}&	&\qw			&\qw			&\ctrl{4}		&\qw			&\ctrl{4}		&\qw			&\qw			&\qw			&\ctrl{4}		&\qw			&\ctrl{4}		&\qw			&\qw		&\rstick{}\\
\lstick{}&	&\qw			&\ctrl{4}		&\qw			&\qw			&\qw			&\ctrl{4}		&\qw			&\ctrl{4}		&\qw			&\qw			&\qw			&\ctrl{4}		&\qw		&\rstick{}\\
\lstick{}&	&\ctrl{4}		&\qw			&\qw			&\qw			&\qw			&\qw			&\ctrl{4}		&\qw			&\qw			&\qw			&\qw			&\qw			&\qw		&\rstick{}\\
\lstick{}&	&\qw			&\qw			&\ctrl{1}		&\targ		&\ctrl{1}		&\qw			&\qw			&\qw			&\ctrl{1}		&\targ		&\ctrl{1}		&\qw			&\qw		&\rstick{}\\
\lstick{}&	&\qw			&\ctrl{1}		&\targ		&\qw			&\targ		&\ctrl{1}		&\qw			&\ctrl{1}		&\targ		&\qw			&\targ		&\ctrl{1}		&\qw		&\rstick{}\\
\lstick{}&	&\ctrl{1}		&\targ		&\qw			&\qw			&\qw			&\targ		&\ctrl{1}		&\targ		&\qw			&\qw			&\qw			&\targ		&\qw		&\rstick{}\\
\lstick{}&	&\targ		&\qw			&\qw			&\qw			&\qw			&\qw			&\targ		&\qw			&\qw			&\qw			&\qw			&\qw			&\qw		&\rstick{}
}
\end{tabular}
\\[1ex]
\\
DR2   & \begin{tabular}[c]{@{}c@{}} if $w \geq 5$ and $\vert C \vert >\lceil w/2 \rceil$, then \\$CNOT(C; t)$ is decomposed into\\ four $CNOT$ gates with $\lceil w/2 \rceil$ \\and $w-1-\lceil w/2\rceil$ control qubits 
 \end{tabular}& 
\begin{tabular}{ccccc}
\Qcircuit @C=0.01em @R=0.5em @!R{
\lstick{}&	&\ctrl{8}		&\qw		&\rstick{}\\
\lstick{}&	&\ctrl{7}		&\qw		&\rstick{}\\
\lstick{}&	&\ctrl{6}		&\qw		&\rstick{}\\
\lstick{}&	&\ctrl{5}		&\qw		&\rstick{}\\
\lstick{}&	&\ctrl{4}		&\qw		&\rstick{}\\
\lstick{}&	&\ctrl{3}		&\qw		&\rstick{}\\
\lstick{}&	&\ctrl{2}		&\qw		&\rstick{}\\
\lstick{}&	&\qw			&\qw		&\rstick{}\\
\lstick{}&	&\targ		&\qw		&\rstick{}
}
&  \raisebox{-15ex}{  $ \equiv $}   &
\Qcircuit @C=0.01em @R=0.5em @!R{
\lstick{}&	&\ctrl{7}		&\qw			&\ctrl{7}		&\qw			&\qw		&\rstick{}\\
\lstick{}&	&\ctrl{6}		&\qw			&\ctrl{6}		&\qw			&\qw		&\rstick{}\\
\lstick{}&	&\ctrl{5}		&\qw			&\ctrl{5}		&\qw			&\qw		&\rstick{}\\
\lstick{}&	&\ctrl{4}		&\qw			&\ctrl{4}		&\qw			&\qw		&\rstick{}\\
\lstick{}&	&\ctrl{3}		&\qw			&\ctrl{3}		&\qw			&\qw		&\rstick{}\\
\lstick{}&	&\qw			&\ctrl{3}		&\qw			&\ctrl{3}		&\qw		&\rstick{}\\
\lstick{}&	&\qw			&\ctrl{2}		&\qw			&\ctrl{2}		&\qw		&\rstick{}\\
\lstick{}&	&\targ		&\ctrl{1}		&\targ		&\ctrl{1}		&\qw		&\rstick{}\\
\lstick{}&	&\qw			&\targ		&\qw			&\targ		&\qw		&\rstick{}
}
&  \raisebox{-15ex}{$ \equiv $ }   &
\Qcircuit @C=0.01em @R=0.5em @!R{
\lstick{}&	&\qw			&\ctrl{7}		&\qw			&\ctrl{7}		&\qw		&\rstick{}\\
\lstick{}&	&\qw			&\ctrl{6}		&\qw			&\ctrl{6}		&\qw		&\rstick{}\\
\lstick{}&	&\qw			&\ctrl{5}		&\qw			&\ctrl{5}		&\qw		&\rstick{}\\
\lstick{}&	&\qw			&\ctrl{4}		&\qw			&\ctrl{4}		&\qw		&\rstick{}\\
\lstick{}&	&\qw			&\ctrl{3}		&\qw			&\ctrl{3}		&\qw		&\rstick{}\\
\lstick{}&	&\ctrl{3}		&\qw			&\ctrl{3}		&\qw			&\qw		&\rstick{}\\
\lstick{}&	&\ctrl{2}		&\qw			&\ctrl{2}		&\qw			&\qw		&\rstick{}\\
\lstick{}&	&\ctrl{1}		&\targ		&\ctrl{1}		&\targ		&\qw		&\rstick{}\\
\lstick{}&	&\targ		&\qw			&\targ		&\qw			&\qw		&\rstick{}
}
\end{tabular}
\\[1ex]
\\
DR3   & \begin{tabular}[c]{@{}c@{}} if $\vert C \vert > 2$ and $\vert C \vert = w-1$, then \\the auxiliary line will be added \\to $CNOT$ gate and DR2 will \\be applied to decompose it
 \end{tabular}& 
 \begin{tabular}{ccccccc}
\Qcircuit @C=0.01em @R=0.6em @!R{
\lstick{}&	&\ctrl{3}		&\qw		&\rstick{}\\
\lstick{}&	&\ctrl{2}		&\qw		&\rstick{}\\
\lstick{}&	&\ctrl{1}		&\qw		&\rstick{}\\
\lstick{}&	&\targ		&\qw		&\rstick{}
}
&  \raisebox{-6ex}{$ \Rightarrow $ }   &
\Qcircuit @C=0.01em @R=0.3em @!R{
\lstick{}&	&\ctrl{4}		&\qw		&\rstick{}\\
\lstick{}&	&\ctrl{3}		&\qw		&\rstick{}\\
\lstick{}&	&\ctrl{2}		&\qw		&\rstick{}\\
\lstick{}&	&\qw			&\qw		&\rstick{}\\
\lstick{}&	&\targ		&\qw		&\rstick{}
}
&  \raisebox{-6ex}{$ \Rightarrow $ }   &
\Qcircuit @C=0.01em @R=0.3em @!R{
\lstick{}&	&\qw			&\ctrl{3}		&\qw			&\ctrl{3}		&\qw		&\rstick{}\\
\lstick{}&	&\qw			&\ctrl{2}		&\qw			&\ctrl{2}		&\qw		&\rstick{}\\
\lstick{}&	&\ctrl{2}		&\qw			&\ctrl{2}		&\qw			&\qw		&\rstick{}\\
\lstick{}&	&\ctrl{1}		&\targ		&\ctrl{1}		&\targ		&\qw		&\rstick{}\\
\lstick{}&	&\targ		&\qw			&\targ		&\qw			&\qw		&\rstick{}
}
 \end{tabular}
 \\[1ex]
\\
\hline                       
\end{tabular}
\caption{The basic decomposition rules \cite{9}.}
\label{decomposition rules}
\end{table}

\begin{Rule}
If two  gates $Controlled-V$ and $CNOT$ are adjacent, then these two gates can be merged into $Controlled-V\dagger$ gate as shown in Fig.~\ref{optimization rules fig}(b).
\end{Rule} 

\begin{Rule}
If two  gates $Controlled-V\dagger$ and $CNOT$ are adjacent, then these two gates can be merged into $Controlled-V$ gate as shown in Fig.~\ref{optimization rules fig}(c).
\end{Rule} 

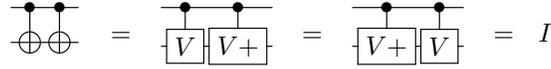
\begin{figure}[!ht]
\centering
\small\addtolength{\tabcolsep}{-3pt}
\begin{tabular}{ccccccc}
\Qcircuit @C=0.3em @R=0.5em @!R{
\lstick{}&	&\ctrl{1}		&\ctrl{1}		&\qw		&\rstick{}\\
\lstick{}&	&\targ		&\targ		&\qw		&\rstick{}
}
&   \raisebox{-3ex}{ $=$ }  &
\Qcircuit @C=0.2em @R=0.1em @!R{
\lstick{}&	&\ctrl{1}			&\ctrl{1}			&\qw		&\rstick{}\\
\lstick{}&	&\gate{V}			&\gate{V+}		&\qw		&\rstick{}
}
&   \raisebox{-3ex}{ $=$ }  &
\Qcircuit @C=0.2em @R=0.1em @!R{
\lstick{}&	&\ctrl{1}			&\ctrl{1}			&\qw		&\rstick{}\\
\lstick{}&	&\gate{V+}		&\gate{V}			&\qw		&\rstick{}
}
&   \raisebox{-3ex}{ $=$ }  &
\raisebox{-3ex}{$I$}
\end{tabular}
\caption{Delete Rule}
\label{delete rules fig}
\end{figure}

\begin{Rule}
If two gates $Controlled-V$ and $Controlled-V\dagger$ or two $CNOT$ gates are adjacent, then these gates can be removed from the quantum circuit as shown in Fig.~\ref{delete rules fig}.
\end{Rule} 

\begin{Rule}
For any two $NCV$ gates $g_{1}(c_{1}, t_{1})$ and $g_{2}(c_{2}, t_{2})$ \\ where $c_{1}, c_{2}, t_{1}$ and $t_{2}$ are the two controls and targets of $g_{1}$ and $g_{2}$ respectively, if $c_{1} \neq t_{2}$ and $t_{2}\neq c_{2}$, then gates can be exchanged. 
\end{Rule}

\begin{algorithm}[!ht]
\caption{Mapping $MCT$ circuit into $NCV$ circuit}\label{alg:cap}
\SetKwInOut{Input}{Input}
\SetKwInOut{Output}{Output}
\SetAlgoNoLine%
        \Input{$MCT$ circuit $RC_{1} = g_{1}, g_{2}, .., g_{r}$ with $w$ qubits}
        \Output{Optimized $NCV$ circuit}
\Begin{\For {$i$ to $r$}
        {$m = \vert C \vert$ of $g_{i}$\\
        \uIf{$m >2$ and $m == w-1$}{
           Apply $DR3$
          }
        \uElseIf{$w \geq 5$ and $3 \leq m \leq \lceil w/2 \rceil$}{
           apply $DR1$
           }
        \uElseIf{$w \geq 5$ and $m >  \lceil w/2 \rceil $} {Apply $DR2$} 
        \Else{Decompose $CNOT$ gate with $\vert C \vert=2$ using different form in Fig.~\ref{Decomposition positive CNOT} and Fig.~\ref{Decomposition negative CNOT}}
         increment $i$       
        }
        Apply simplification rules\\
        Remove the last gates that restore the states of input qubits. 
}
\end{algorithm}

\section{Experimental Results}
\label{sec:result}

The results of the proposed optimization algorithms have been compared with the results of RevLib \cite{ref}, the genetic algorithm $GA$ in \cite{7}, the reorder method in \cite{journals/qip/AhmedYE18} and another methods in \cite{11}, \cite{22}, \cite{25}. The experimental results are displayed in Table~\ref{tab:revlib}, Table~\ref{tab:reorder}, Table~\ref{tab:GA}, Table~\ref{tab:2020v1}, Table~\ref{tab:2020v2} and Table~\ref{tab:2021} where the proposed algorithm achieves a reduction in the quantum cost of synthesized quantum circuits on average compared with other algorithms in literature. 

In Table~\ref{tab:revlib}, a comparison between the synthesized circuits in RevLib \cite{ref} with $MCT$ library and the proposed algorithm is displayed, where the quantum cost $(QC)$ of all functions has been improved compared to RevLib \cite{ref}.
\begin{table}[!ht]
\centering
\small\addtolength{\tabcolsep}{-3.5pt}
\begin{tabular} { c  c  c  c  c c c c c }
 \hline 
Function&\multicolumn{2}{c}{Quantum Cost} & Function&\multicolumn{2}{c}{Quantum Cost} \\  [0.5ex]\cline{2-3} \cline{5-6}

      & RevLib \cite{ref} & The Proposed &     & RevLib \cite{ref} & The Proposed\\ [0.5ex]
      &                   &Algorithm &	&	&Algorithm \\[0.5ex]
\hline
4gt4\_20		& 54		& 45  & sym6\_63	& 777	& 262\\
4gt5\_21		& 21		& 11  & 5ex2\_151	& 141	& 108\\
4gt10\_22		& 34		& 28  & 5ex3\_152	& 79	& 70\\
4gt11\_23		& 7			& 4   & 5xor5\_195	& 7		& 5\\
4gt12\_24		& 41		& 37  & 11cm152a\_130	& 252	&114\\
4gt13\_25		& 15		& 10  & 4mod5\_8		& 9	& 6\\
4sf\_232		& 51		& 19  & 9symml			& 14,193	&693\\
9sym9\_71		& 4368		&   693  &5majority\_176	& 136	& 118 \\ \cline{4-6}
9life			& 6766		& 1843 & Average & 1585.4 & 239.2 \\
\hline 

\end{tabular}
\caption{The Boolean function synthesized in RevLib \cite{ref} vs using the proposed algorithm.}
\label{tab:revlib}
\end{table}

Table~\ref{tab:reorder} displays a comparison between the reorder method \cite{journals/qip/AhmedYE18} and the proposed algorithm, where the quantum cost $(QC)$ of quantum circuits of 88 $\%$ Boolean functions are enhanced. The proposed algorithm didn't achieve any improvement in the quantum cost $(QC)$ of quantum circuits for three functions. For example, all $BFV8\_d3$ Boolean functions family cannot be improved because the degree of these functions is $d=3$ that can be realized by generalized Toffoli gate with three controls that can be decomposed into $14$ elementary gates but its quantum cost $(QC)$ is $13$.

Table~\ref{tab:GA} shows a comparison between the genetic algorithm $GA$ \cite{7} and the proposed optimization algorithms. The proposed algorithm achieves a reduction in the quantum cost (QC) of the quantum circuits for all the Boolean functions except five functions, $It42$, $It52$, $8newill$, $4gt4\_20$ and  $4gt10\_24$. The proposed algorithm didn't improve the quantum cost $(QC)$ of  $It42$, $It52$, $4gt4\_20$ and  $4gt10\_24$ because the number of the product terms that have the same degree is less than three product terms, so the algorithm can't applied. In $8newill$, there is no common factor variables $[v]$ between the $[g][v]$ terms, so the proposed algorithm didn't achieve a reduction in the quantum cost. 

\begin{table}[!ht]
\centering
\small\addtolength{\tabcolsep}{-4.5pt}
\begin{tabular} { c  c  c  c  c c c c c }
 \hline 
Function&\multicolumn{2}{c}{Quantum Cost} & Function&\multicolumn{2}{c}{Quantum Cost} \\  [0.5ex]\cline{2-3} \cline{5-6}

      &  Method \cite{journals/qip/AhmedYE18} & The Proposed &     & Method \cite{journals/qip/AhmedYE18} & The Proposed\\ [0.5ex]
      &											& Algorithm &		&									& Algorithm\\[0.5ex]
\hline
It41	&	26  &	22    & 4gt4\_20	&   51	& 46\\
It42	&	56	&	50	  & 4gt5\_21	&	14	& 11\\
It43	&	28	&	21	  & 4gt10\_22	&	34	& 28\\
It44	&	42	&   41	  & 4gt11\_23	&	5	& 4 \\
It45	&	58	&   47	  & 4gt12\_24	&	43	& 37\\
It51	&	141	&   119   & 4gt13\_25	&	13	& 10\\
It52	&	128	&   103   & 4mod5\_8	&	8	& 6\\
4sf\_232 &	22	&   19	  & sym6\_63	&	269	& 262\\
7rd73f1	&	25	&   14    & 8rd84f1  	&	29	& 16\\
7rd73f2	&	7	&   7     & 8rd84f2	    &	8	& 8\\
7rd73f3	&	277	&  267    & 8rd84f3		&	509	& 85\\
7con1f1	&	121	&  106    & 8rd84f4		&	418	& 408\\
7con1f2	&	61	&  49	  & 5majority\_176	&	138	& 118\\
5ex2\_151	&	129	&108 & 11cm152a\_130	&	119&	114\\
5ex3\_152	&	74	& 70  & 9sym9\_71		&	695	& 693\\
5rd53f1		&	90	&  100& 8newill			&	920 &	927\\
5rd53f2		&	17	&  10 & 5alu\_9	&	24	&22\\ 
8newtag		&	387	& 372 & sym(4,2,4)	&	10	& 7\\
sym(4,2,6)	&	13	& 8   & sym(6,2,16)	&	12	&9\\
sym(4,3,4)	&	17	& 13  & sym(6,3,12)	&	44	&43\\
sym(5,2,8)	&	11	& 8   & sym(6,3,16)	&	71	&70\\
sym(5,2,12)	&	14	& 9 	& BFV8\_d3	&	85	& 87\\
sym(5,3,4)	&	28	& 26	& BFV8\_d3\_24	&	146	& 152 \\
sym(5,3,6)	&	47	& 46	& BFV8\_d3\_28	&	171	& 179 \\
sym(5,3,10)	&	47	& 46	& BFV8\_d3\_32	&	177	& 184\\ \cline{4-6}
2 of 5		&	51	& 51	& Average       &	116.27 & 103.06	\\
\hline 

\end{tabular}
\caption{The reversible Boolean function synthesized with reorder method in \cite{journals/qip/AhmedYE18} vs using the proposed algorithm.}
\label{tab:reorder}
\end{table}

\begin{table}[!ht]
\centering
\small\addtolength{\tabcolsep}{-3pt}
\begin{tabular} { c  c  c  c  c c c c c }
 \hline 
Function&\multicolumn{2}{c}{Quantum Cost} & Function&\multicolumn{2}{c}{Quantum Cost} \\  [0.5ex]\cline{2-3} \cline{5-6}

      & GA \cite{7} & The Proposed &     & GA \cite{7} & The Proposed \\ [0.5ex]
      &				& Algorithm   &		&			& Algorithm\\[0.5ex]	
\hline
It41		& 28 	& 22  & 4gt4\_20		&36 &	46\\
It42		& 42	& 50  & 4gt5\_21		& 14 &	11\\
It43		& 31	& 21  & 4gt10\_22		& 34	&28\\
It44		& 46	& 41  & 4gt11\_23		&5	& 4\\
It45		& 55	& 47  & 4gt10\_24		&34 &	37\\
It51		& 140	& 119 & 4gt10\_25		&13	& 10\\
It52		& 100	& 103 & 4sf\_232		&29	& 19\\
7con1f1		&131	&106  & 5alu\_9		   &36	&22\\
7con1f2		& 67	& 49  & 8newill			&822	&927\\
sf\_276		& 25	& 18  & 7rd73f2			&7	&7\\
sf\_275		& 31	& 18  & 8newtag			& 437	&372\\
8rd84f2		& 8		& 8   & sym6\_145		&595	&262\\
8rd84f3		& 509	& 85  & 4Mod5\_8		&21	&6\\
majority\_239	&128	&118 & cm152a\_212		&281	&114\\ \cline{4-6}
5rd53f2		&50	&10     & Average & 129.48& 92.41\\
\hline 

\end{tabular}
\caption{The reversible Boolean function synthesized with genetic algorithm $GA$ in \cite{7} vs using the proposed algorithm.}
\label{tab:GA}
\end{table}

In Table~\ref{tab:2020v1} and Table~\ref{tab:2020v2}, a comparison between synthesized circuits of the Boolean functions with a single output using the method in \cite{11}, \cite{22} respectively and the proposed algorithm is displayed. The proposed algorithm achieves a reduction in quantum cost of quantum circuits for all functions compared with other methods.  The last row in Table~\ref{tab:2020v1} and Table~\ref{tab:2020v2} displays the average of the quantum cost $QC$ for method \cite{11}, method \cite{22} respectively and the proposed algorithm for each of them.

\begin{table}[!ht]
\centering
\begin{tabular} { c  c  c     }
 \hline 
Function&\multicolumn{2}{c}{Quantum Cost}\\  [0.5ex]\cline{2-3} 

      &  Method \cite{11}  & The Proposed Algorithm \\ [0.5ex]
\hline
4gt5\_75		& 61 	& 11\\
4gt11\_84		& 16	& 4 \\
alu\_v4\_36		& 68	&  22 \\
4gt10\_v1\_81	& 99	& 30\\
4mod5\_v1\_23	& 51	& 6 \\
4gt12\_v1\_89	& 185	& 38\\
4gt13\_v1\_93	& 41	& 1\\ \cline{1-3}
Average			& 74.43	&  16 	\\
\hline 

\end{tabular}
\caption{The reversible Boolean function synthesized with method in \cite{11} vs using the proposed algorithm.}
\label{tab:2020v1}
\end{table}

\begin{table}[!ht]
\centering
\begin{tabular} { c  c  c     }
 \hline 
Function&\multicolumn{2}{c}{Quantum Cost}\\  [0.5ex]\cline{2-3} 

      &  Method \cite{22} & The Proposed Algorithm \\ [0.5ex]
\hline
4gt5\_75		& 48	& 11\\
4gt11\_84		& 11	& 4 \\
4gt11\_83		& 12	& 4 \\
xor5\_254		& 9		& 5\\
alu\_v4\_36		& 64	&  22 \\
4mod5\_v1\_23	& 17	& 6 \\
4mod5\_v1\_22   & 10 	& 6 \\
4gt11\_v1\_85	& 12	& 5 \\
4gt13\_v1\_93	& 38	& 1\\ \cline{1-3}
Average			& 24.56	&  7.11	\\
\hline 

\end{tabular}
\caption{The reversible Boolean function synthesized with method in \cite{22} vs using the proposed algorithm.}
\label{tab:2020v2}
\end{table}

Table~\ref{tab:2021} shows a comparison between the method \cite{25} and the proposed algorithm for all functions with a single output. The $rd53$ and $rd73$ functions have three outputs and the $rd84$ has four outputs. The proposed algorithm is applied on these functions for each output individually. This means that the quantum cost of $rd73$ circuit, for example, is the sum of quantum cost of $rd73f1$, $rd73f2$ and $rd73f3$. The proposed algorithm achieves a reduction in quantum cost of quantum circuits on average.   

\begin{table}[!ht]
\centering
\begin{tabular} { c  c  c  }
 \hline 
Function&\multicolumn{2}{c}{Quantum Cost}\\  [0.5ex]\cline{2-3} 

      &  Method \cite{25} &  The Proposed Algorithm \\ [0.5ex]
\hline
4mod5       &  17   & 6\\
alu         & 29	& 22\\
rd53		& 117	& 115\\
rd73		& 357	& 288 \\
rd84		& 588	& 516 \\
sym6		& 114	& 262 \\
sym9		& 880	& 693 \\
xor5		& 4 	& 5   \\
majority	& 38	& 118	\\
parity		& 15	& 16 \\ \cline{1-3}
Average		& 215.9	& 204.1			\\
\hline 
\end{tabular}
\caption{The reversible Boolean function synthesized with method in \cite{25} vs using the proposed algorithm.}
\label{tab:2021}
\end{table}

\section{Discussion} 
\label{sec:Discussion}
\begin{figure}[!ht]
\centering
\begin{tabular}{ccc}
\includegraphics[scale=0.5]{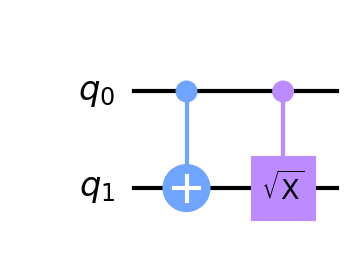}
&  \raisebox{5ex}{\,\,\,\,\,\ $\equiv $ }   &
\includegraphics[scale=0.5]{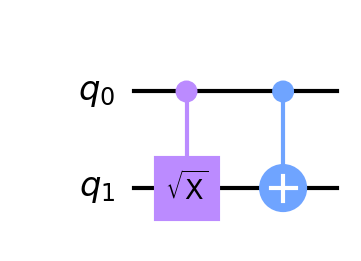}
\end{tabular}
\caption{Two various realization of the $Controlled-V\dagger$ gate on IBM quantum computer.}
\label{vdash}
\end{figure}

\begin{figure}[!ht]
\centerline{\includegraphics[scale=0.5]{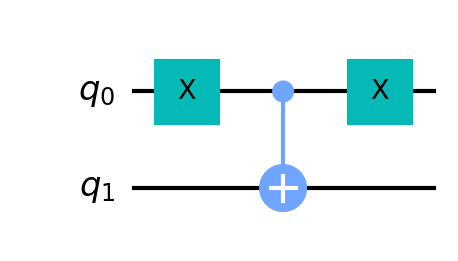}}
\caption{Realization of the $NOT$ gate on IBM quantum computer.}
\label{not gate}
\end{figure}

IBM quantum computers \cite{ibm} are quantum computers that uses universal gates with one-qubit and two-qubit only to synthesize a given quantum circuit and so the quantum gates for more than two qubits can't be implemented. So, the main step to implement a quantum circuit of $MCT$ gates on IBM quantum computer is to decompose $MCT$ gates into one-qubit gates and two-qubit gates. The proposed algorithm decomposes the $MCT$ gates into the $NCV$ quantum gates, one-qubit and two-qubit gates, which will help in implementation of quantum circuit for a given Boolean function expressed in $PPRM$ on IBM quantum computer. The proposed algorithm proposes a special algebraic form for a given Boolean function to minimize the quantum cost for equivalent quantum circuits. The synthesized quantum circuits achieve a reduction in number of gates, volume of quantum circuit and thus time of computation. IBM quantum computer supports finite quantum gates, however, another quantum gates can't be supported such as $Controlled-V\dagger$ gate and controlled quantum gate with negative condition on the control qubit. The $Controlled-V\dagger$ gate can be replaced by two quantum gates $Controlled-V$ gate followed by $CNOT$ gate or $CNOT$ gate followed by $Controlled-V$ gate to be implemented on IBM quantum computer as shown in Fig.~\ref{vdash}. To implement a quantum gate with a control qubit with negative condition on IBM quantum computer, the control qubit with negative condition can be replaced by control qubit with positive condition after adding a $NOT$ gate before and after the control qubit with positive condition \cite{Arabzadeh:2010:ROR:1899721.1899916} as shown in Fig.~\ref{not gate}. 

\begin{example}
Consider the following Boolean function $f$ of degree 4 for the $4gt4\_20$ quantum circuit after applying the reorder method \cite{journals/qip/AhmedYE18},
\begin{equation}
f(x_{1}, x_{2}, x_{3}, x_{4})= x_{1}\oplus x_{1}x_{2}x_{3}\bar{x_{4}}\oplus (x_{2}x_{4})(x_{1}\oplus x_{3})\oplus x_{2}(x_{3}\oplus x_{4})
\end{equation}

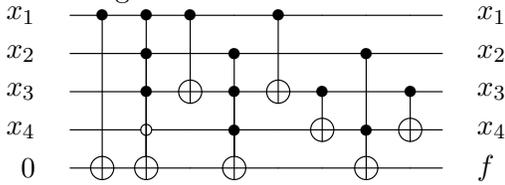
\begin{figure}[!ht]
\centering
\Qcircuit @C=0.7em @R=0.5em @!R{
\lstick{x_{1}}&	&\ctrl{4}		&\ctrl{4}		&\ctrl{2}		&\qw			&\ctrl{2}		&\qw			&\qw			&\qw			&\qw		&\rstick{x_{1}}\\
\lstick{x_{2}}&	&\qw			&\ctrl{3}		&\qw			&\ctrl{3}		&\qw			&\qw			&\ctrl{3}		&\qw			&\qw		&\rstick{x_{2}}\\
\lstick{x_{3}}&	&\qw			&\ctrl{2}		&\targ		&\ctrl{2}		&\targ		&\ctrl{1}		&\qw			&\ctrl{1}		&\qw		&\rstick{x_{3}}\\
\lstick{x_{4}}&	&\qw			&\ctrlo{1}		&\qw			&\ctrl{1}		&\qw			&\targ		&\ctrl{1}		&\targ		&\qw		&\rstick{x_{4}}\\
\lstick{0}&	&\targ		&\targ		&\qw			&\targ		&\qw			&\qw			&\targ		&\qw			&\qw		&\rstick{f}
}
\caption{the $4gt4\_20$ quantum circuit after applying the reorder method \cite{journals/qip/AhmedYE18}}
\label{fn_reorder1_ex}
\end{figure}

Now, the $d_{term}$ can be calculated for each term in the Boolean function $f$ according to Table.~\ref{degree term table} as follows:
\begin{enumerate}
\item The $d_{term}$ for two product terms $x_{1}$ and $x_{1}x_{2}x_{3}\bar{x_{4}}$ is 1 and 4 respectively as Case 1. 
\item The $d_{term}$ for $(x_{2}x_{4})(x_{1}\oplus x_{3})$ term is 3 as Case 3.
\item The $d_{term}$ for $ x_{2}(x_{3}\oplus x_{4})$ term is 2 as Case 4.
\end{enumerate}
Then find the term with the maximum $d_{term}$ ,$x_{1}x_{2}x_{3}\bar{x_{4}}$ term, and rewrite the Boolean function $f$ to be,
\begin{equation}
f(x_{1}, x_{2}, x_{3}, x_{4})= x_{1}\oplus (x_{2}x_{4})(x_{1}\oplus x_{3})\oplus x_{2}(x_{3}\oplus x_{4})\oplus x_{1}x_{2}x_{3}\bar{x_{4}}
\label{circuit after rewrite}
\end{equation}
Then apply basic decomposition rules and simplification rules to map the $MCT$ circuit into the $NCV$ circuit as the second stage of the proposed algorithm. The $4gt4\_20$ quantum circuit has three $CNOT$ gates with more than one control that must be decomposed. So, the extra line can be added to apply decomposition rule $DR3$ on the last gate as shown in Fig.~\ref{fn_ex_extraline}. 

\begin{figure}[!htb]
\centering
\small\addtolength{\tabcolsep}{-3pt}
\begin{tabular}{ccc}
\Qcircuit @C=0.7em @R=0.9em @!R{
\lstick{x_{1}}&	&\ctrl{4}		&\ctrl{2}		&\qw			&\ctrl{2}		&\qw			&\qw			&\qw			&\ctrl{4}		&\qw		&\rstick{}\\
\lstick{x_{2}}&	&\qw			&\qw			&\ctrl{3}		&\qw			&\qw			&\ctrl{3}		&\qw			&\ctrl{3}		&\qw		&\rstick{}\\
\lstick{x_{3}}&	&\qw			&\targ		&\ctrl{2}		&\targ		&\ctrl{1}		&\qw			&\ctrl{1}		&\ctrl{2}		&\qw		&\rstick{}\\
\lstick{x_{4}}&	&\qw			&\qw			&\ctrl{1}		&\qw			&\targ		&\ctrl{1}		&\targ		&\ctrlo{1}		&\qw		&\rstick{}\\
\lstick{0}&	&\targ		&\qw			&\targ		&\qw			&\qw			&\targ		&\qw			&\targ		&\qw		&\rstick{}
}
&  \raisebox{-10ex}{  $ \Rightarrow $ \,\,\,\,\,\,\,\,}   &
\Qcircuit @C=0.5em @R=0.5em @!R{
\lstick{x_{1}}&	&\ctrl{5}		&\ctrl{2}		&\qw			&\ctrl{2}		&\qw			&\qw			&\qw			&\ctrl{5}		&\qw		&\rstick{gr}\\
\lstick{x_{2}}&	&\qw			&\qw			&\ctrl{4}		&\qw			&\qw			&\ctrl{4}		&\qw			&\ctrl{4}		&\qw		&\rstick{gr}\\
\lstick{x_{3}}&	&\qw			&\targ		&\ctrl{3}		&\targ		&\ctrl{1}		&\qw			&\ctrl{1}		&\ctrl{3}		&\qw		&\rstick{gr}\\
\lstick{x_{4}}&	&\qw			&\qw			&\ctrl{2}		&\qw			&\targ		&\ctrl{2}		&\targ		&\ctrlo{2}		&\qw		&\rstick{gr}\\
\lstick{L_{1}}&	&\qw			&\qw			&\qw			&\qw			&\qw			&\qw			&\qw			&\qw			&\qw		&\rstick{gr}\\
\lstick{0}&	&\targ		&\qw			&\targ		&\qw			&\qw			&\targ		&\qw			&\targ		&\qw		&\rstick{f}
}
\end{tabular}
\caption{The $4gt4\_20$ quantum circuit in Eq.~\ref{circuit after rewrite} after adding extra line.}
\label{fn_ex_extraline}
\end{figure}
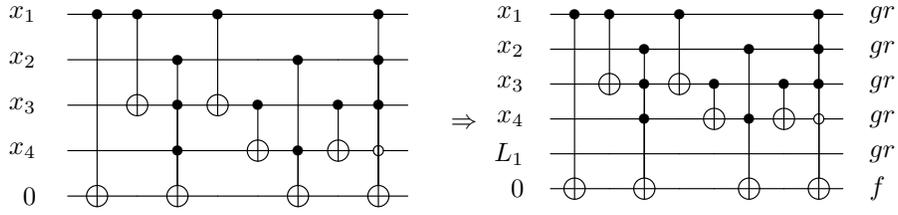

Now, we analyze the way that the proposed algorithm decreases the quantum cost of the $4gt4\_20$ quantum circuit as shown in Fig.~\ref{fig:analysis circuit}. $DR2$ can be applied on the last gate of the $4gt4\_20$ quantum circuit in Fig.~\ref{fig:a} to get a quantum circuit as shown in Fig.~\ref{fig:b} with quantum cost 52. The quantum cost can be decreased to be 46 by removing the gates that have no effect on the result as shown in Fig.~\ref{fig:c}. Finally, apply the decomposition rules in Table.~\ref{decomposition rules} to decompose the $4gt4\_20$ circuit and then the simplification rules to get $4gt4\_20$ circuit as shown in Fig.~\ref{fn_ex_decompose_ncv} with quantum cost 47.

\begin{figure}[!htb]
\centering
\begin{subfigure}{0.4\textwidth}
    \[
\Qcircuit @C=0.2em @R=0.2em @!R{
\lstick{x_{1}}&	&\ctrl{5}		&\ctrl{2}		&\qw			&\ctrl{2}		&\qw			&\qw			&\qw			&\ctrl{5}		&\qw		&\rstick{gr}\\
\lstick{x_{2}}&	&\qw			&\qw			&\ctrl{4}		&\qw			&\qw			&\ctrl{4}		&\qw			&\ctrl{4}		&\qw		&\rstick{gr}\\
\lstick{x_{3}}&	&\qw			&\targ		&\ctrl{3}		&\targ		&\ctrl{1}		&\qw			&\ctrl{1}		&\ctrl{3}		&\qw		&\rstick{gr}\\
\lstick{x_{4}}&	&\qw			&\qw			&\ctrl{2}		&\qw			&\targ		&\ctrl{2}		&\targ		&\ctrlo{2}		&\qw		&\rstick{gr}\\
\lstick{L_{1}}&	&\qw			&\qw			&\qw			&\qw			&\qw			&\qw			&\qw			&\qw			&\qw		&\rstick{gr}\\
\lstick{0}&	&\targ		&\qw			&\targ		&\qw			&\qw			&\targ		&\qw			&\targ		&\qw		&\rstick{f}
}
\]
    \caption{The $4gt4\_20$ circuit.}
    \label{fig:a}
\end{subfigure}
\hfill
\begin{subfigure}{0.5\textwidth}
  \[
\Qcircuit @C=0.2em @R=0.2em @!R{
\lstick{x_{1}}&	&\ctrl{5}		&\ctrl{2}		&\qw			&\ctrl{2}		&\qw			&\qw			&\qw			&\qw			&\ctrl{4}		&\qw			&\ctrl{4}		&\qw		&\rstick{gr}\\
\lstick{x_{2}}&	&\qw			&\qw			&\ctrl{4}		&\qw			&\qw			&\ctrl{4}		&\qw			&\qw			&\ctrl{3}		&\qw			&\ctrl{3}		&\qw		&\rstick{gr}\\
\lstick{x_{3}}&	&\qw			&\targ		&\ctrl{3}		&\targ		&\ctrl{1}		&\qw			&\ctrl{1}		&\qw			&\ctrl{2}		&\qw			&\ctrl{2}		&\qw		&\rstick{gr}\\
\lstick{x_{4}}&	&\qw			&\qw			&\ctrl{2}		&\qw			&\targ		&\ctrl{2}		&\targ		&\ctrlo{2}		&\qw			&\ctrlo{2}		&\qw			&\qw		&\rstick{gr}\\
\lstick{L_{1}}&	&\qw			&\qw			&\qw			&\qw			&\qw			&\qw			&\qw			&\ctrl{1}		&\targ		&\ctrl{1}		&\targ		&\qw		&\rstick{gr}\\
\lstick{0}&	&\targ		&\qw			&\targ		&\qw			&\qw			&\targ		&\qw			&\targ		&\qw			&\targ		&\qw			&\qw		&\rstick{f}
}
\]
    \caption{The $4gt4\_20$ circuit with quantum cost 52 after applying $DR2$ on the last gate.}
    \label{fig:b}
\end{subfigure}
\hfill
\begin{subfigure}{0.6\textwidth}
\[
\Qcircuit @C=0.2em @R=0.2em @!R{
\lstick{x_{1}}&	&\ctrl{5}		&\ctrl{2}		&\qw			&\ctrl{2}		&\qw			&\qw			&\qw			&\qw			&\ctrl{4}		&\qw			&\qw		&\rstick{gr}\\
\lstick{x_{2}}&	&\qw			&\qw			&\ctrl{4}		&\qw			&\qw			&\ctrl{4}		&\qw			&\qw			&\ctrl{3}		&\qw			&\qw		&\rstick{gr}\\
\lstick{x_{3}}&	&\qw			&\targ		&\ctrl{3}		&\targ		&\ctrl{1}		&\qw			&\ctrl{1}		&\qw			&\ctrl{2}		&\qw			&\qw		&\rstick{gr}\\
\lstick{x_{4}}&	&\qw			&\qw			&\ctrl{2}		&\qw			&\targ		&\ctrl{2}		&\targ		&\ctrlo{2}		&\qw			&\ctrlo{2}		&\qw		&\rstick{gr}\\
\lstick{L_{1}}&	&\qw			&\qw			&\qw			&\qw			&\qw			&\qw			&\qw			&\ctrl{1}		&\targ		&\ctrl{1}		&\qw		&\rstick{gr}\\
\lstick{0}&	&\targ		&\qw			&\targ		&\qw			&\qw			&\targ		&\qw			&\targ		&\qw			&\targ		&\qw		&\rstick{f}
}
\]
    \caption{The $4gt4\_20$ circuit with quantum cost 46 after removing the gate that has not effect on the result.}
    \label{fig:c}
\end{subfigure}
        
\caption{The $4gt4\_20$ circuit with minimum quantum cost.}
\label{fig:analysis circuit}
\end{figure}
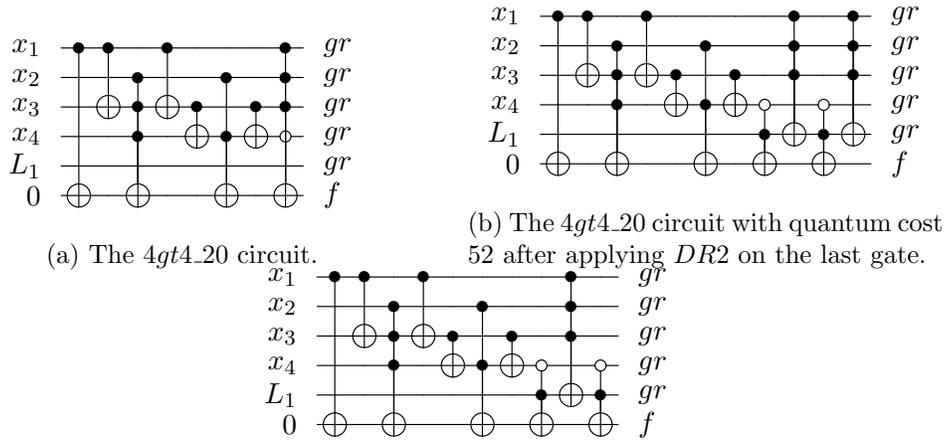

\begin{figure}[!ht]
\centering
\small\addtolength{\tabcolsep}{-10pt}
\begin{tabular}{c}
\Qcircuit @C=0.2em @R=0.5em @!R{
\lstick{x_{1}}&	&\ctrl{2}		&\ctrl{1}			&\targ		&\ctrl{1}			&\gate{V}\qwx[1]	&\qw			&\gate{V+}\qwx[1]	&\gate{V}\qwx[1]	&\ctrl{1}			&\targ		&\ctrl{1}			&\gate{V+}\qwx[1]	&\gate{V}\qwx[1]	&\qw			&\gate{V+}\qwx[1]	&\ctrl{2}		&\qw			&\qw				&\qw			&\qw				&\qw			&\qw				&\qw		&\rstick{}\\
\lstick{x_{2}}&	&\qw			&\qw\qwx[1]		&\qw			&\qw\qwx[1]		&\qw\qwx[1]		&\ctrl{1}		&\qw\qwx[1]		&\ctrl{0}			&\qw\qwx[1]		&\qw			&\qw\qwx[1]		&\ctrl{0}			&\qw\qwx[1]		&\ctrl{1}		&\qw\qwx[1]		&\qw			&\qw			&\qw				&\ctrl{2}		&\qw				&\ctrl{2}		&\ctrl{1}			&\qw		&\rstick{}\\
\lstick{x_{3}}&	&\targ		&\qw\qwx[1]		&\qw			&\qw\qwx[1]		&\ctrl{0}			&\targ		&\ctrl{0}			&\qw				&\qw\qwx[1]		&\qw			&\qw\qwx[1]		&\qw				&\ctrl{0}			&\targ		&\ctrl{0}			&\targ		&\ctrl{1}		&\qw				&\qw			&\qw				&\qw			&\qw\qwx[1]		&\qw		&\rstick{}\\
\lstick{x_{4}}&	&\qw			&\qw\qwx[1]		&\ctrl{-3}		&\qw\qwx[1]		&\qw				&\qw			&\qw				&\qw				&\qw\qwx[1]		&\ctrl{-3}		&\qw\qwx[1]		&\qw				&\qw				&\qw			&\qw				&\qw			&\targ		&\ctrl{1}			&\targ		&\ctrl{1}			&\targ		&\qw\qwx[1]		&\qw		&\rstick{}\\
\lstick{L_{1}}&	&\qw			&\qw\qwx[1]		&\qw			&\qw\qwx[1]		&\qw				&\qw			&\qw				&\qw				&\qw\qwx[1]		&\qw			&\qw\qwx[1]		&\qw				&\qw				&\qw			&\qw				&\qw			&\qw			&\qw\qwx[1]		&\qw			&\qw\qwx[1]		&\qw			&\qw\qwx[1]		&\qw		&\rstick{}\\
\lstick{0}&	&\qw			&\gate{V+}		&\qw			&\gate{V+}		&\qw				&\qw			&\qw				&\qw				&\gate{V}			&\qw			&\gate{V+}		&\qw				&\qw				&\qw			&\qw				&\qw			&\qw			&\gate{V}			&\qw			&\gate{V+}		&\qw			&\gate{V}			&\qw		&\rstick{}
}
\\
\\
\Qcircuit @C=0.01em @R=0.5em @!R{
\lstick{x_{1}}&	&\qw			&\qw				&\qw			&\qw				&\qw				&\qw				&\qw			&\qw				&\qw				&\ctrl{1}		&\qw				&\ctrl{1}			&\qw				&\qw			&\qw				&\ctrl{1}			&\qw				&\ctrl{1}		&\qw				&\qw				&\qw			&\qw				&\qw				&\qw		&\rstick{gr}\\
\lstick{x_{2}}&	&\qw			&\qw				&\qw			&\qw				&\qw				&\qw				&\qw			&\qw				&\ctrl{1}			&\targ		&\ctrl{1}			&\qw\qwx[1]		&\qw				&\qw			&\qw				&\qw\qwx[1]		&\ctrl{1}			&\targ		&\ctrl{1}			&\qw				&\qw			&\qw				&\qw				&\qw		&\rstick{gr}\\
\lstick{x_{3}}&	&\ctrl{1}		&\qw				&\qw			&\qw				&\qw				&\qw				&\ctrl{1}		&\qw				&\qw\qwx[1]		&\qw			&\qw\qwx[1]		&\qw\qwx[1]		&\qw				&\ctrl{1}		&\qw				&\qw\qwx[1]		&\qw\qwx[1]		&\qw			&\qw\qwx[1]		&\qw				&\qw			&\qw				&\qw				&\qw		&\rstick{gr}\\
\lstick{x_{4}}&	&\targ		&\qw				&\ctrl{1}		&\qw				&\ctrl{1}			&\ctrl{1}			&\targ		&\ctrl{1}			&\gate{V}			&\qw			&\gate{V+}		&\gate{V}			&\ctrl{1}			&\targ		&\ctrl{1}			&\gate{V+}		&\gate{V}			&\qw			&\gate{V+}		&\qw				&\ctrl{1}		&\qw				&\ctrl{1}			&\qw		&\rstick{gr}\\
\lstick{L_{1}}&	&\qw			&\ctrl{1}			&\targ		&\ctrl{1}			&\qw\qwx[1]		&\gate{V+}		&\qw			&\gate{V+}		&\qw				&\qw			&\qw				&\qw				&\gate{V}			&\qw			&\gate{V+}		&\qw				&\qw				&\qw			&\qw				&\ctrl{1}			&\targ		&\ctrl{1}			&\qw\qwx[1]		&\qw		&\rstick{gr}\\
\lstick{0}&	&\qw			&\gate{V}			&\qw			&\gate{V}			&\gate{V+}		&\qw				&\qw			&\qw				&\qw				&\qw			&\qw				&\qw				&\qw				&\qw			&\qw				&\qw				&\qw				&\qw			&\qw				&\gate{V}			&\qw			&\gate{V}			&\gate{V+}		&\qw		&\rstick{f}
}
\end{tabular}
\caption{The $4gt4\_20$ quantum circuit with $NCV$ gates with quantum cost 46.}
\label{fn_ex_decompose_ncv}
\end{figure}
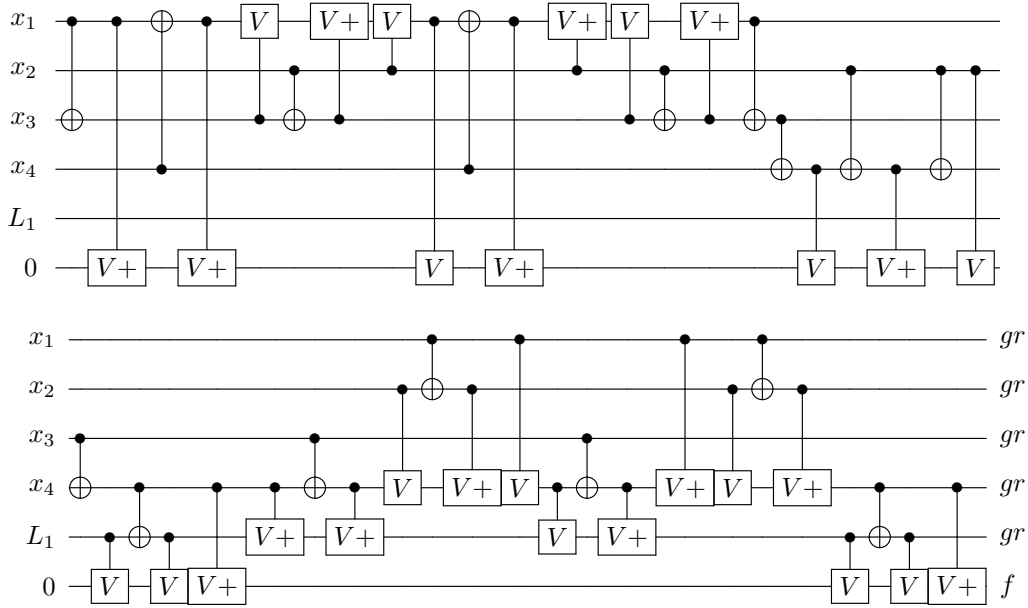

The $4gt4\_20$ circuit with the $MCT$ gates and the $NCV$ gates can be implemented on IBM Q device as shown in Fig.~\ref{mct on ibm} and Fig.~\ref{ncv on ibm} respectively. 
\begin{figure}[!ht]
\centerline{\includegraphics[scale=0.4]{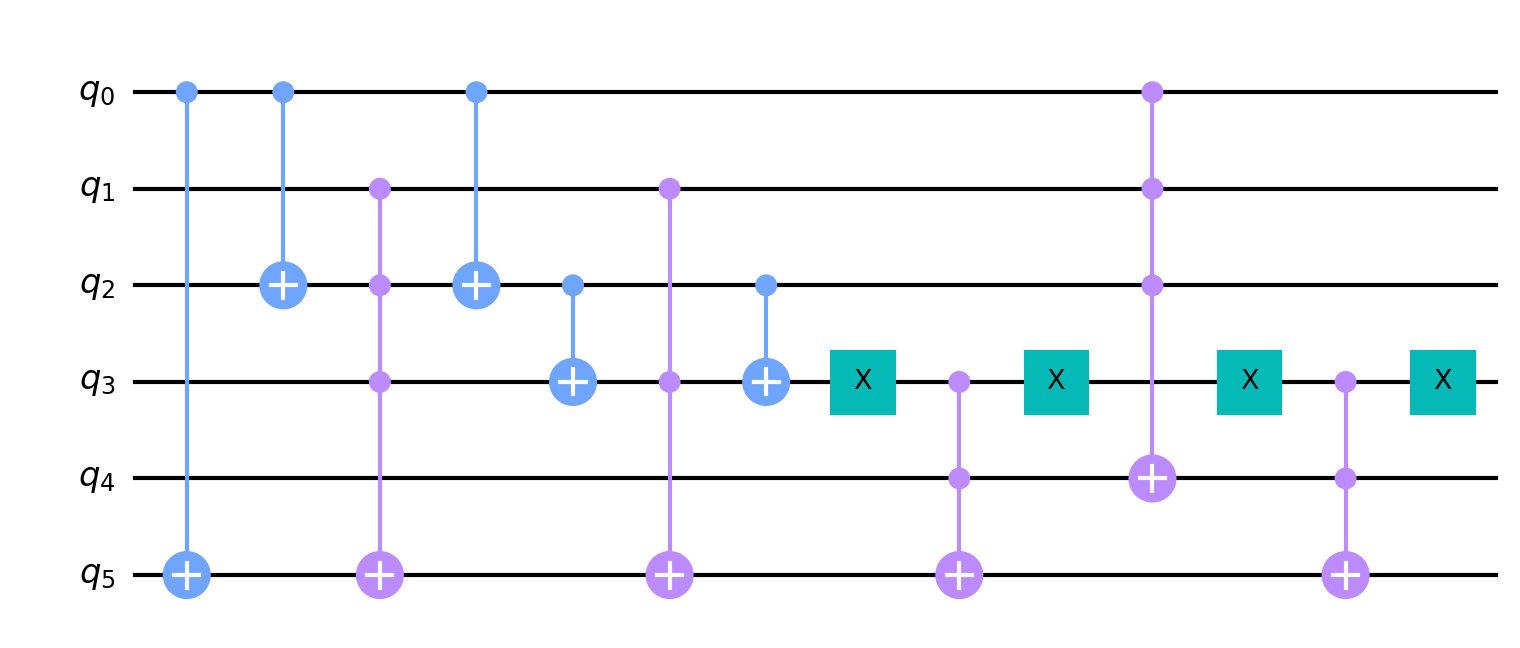}}
\caption{The $4gt4\_20$ quantum circuit with $MCT$ gates using IBM.}
\label{mct on ibm}
\end{figure}

\begin{figure}[!ht]
\centerline{\includegraphics[scale=0.2]{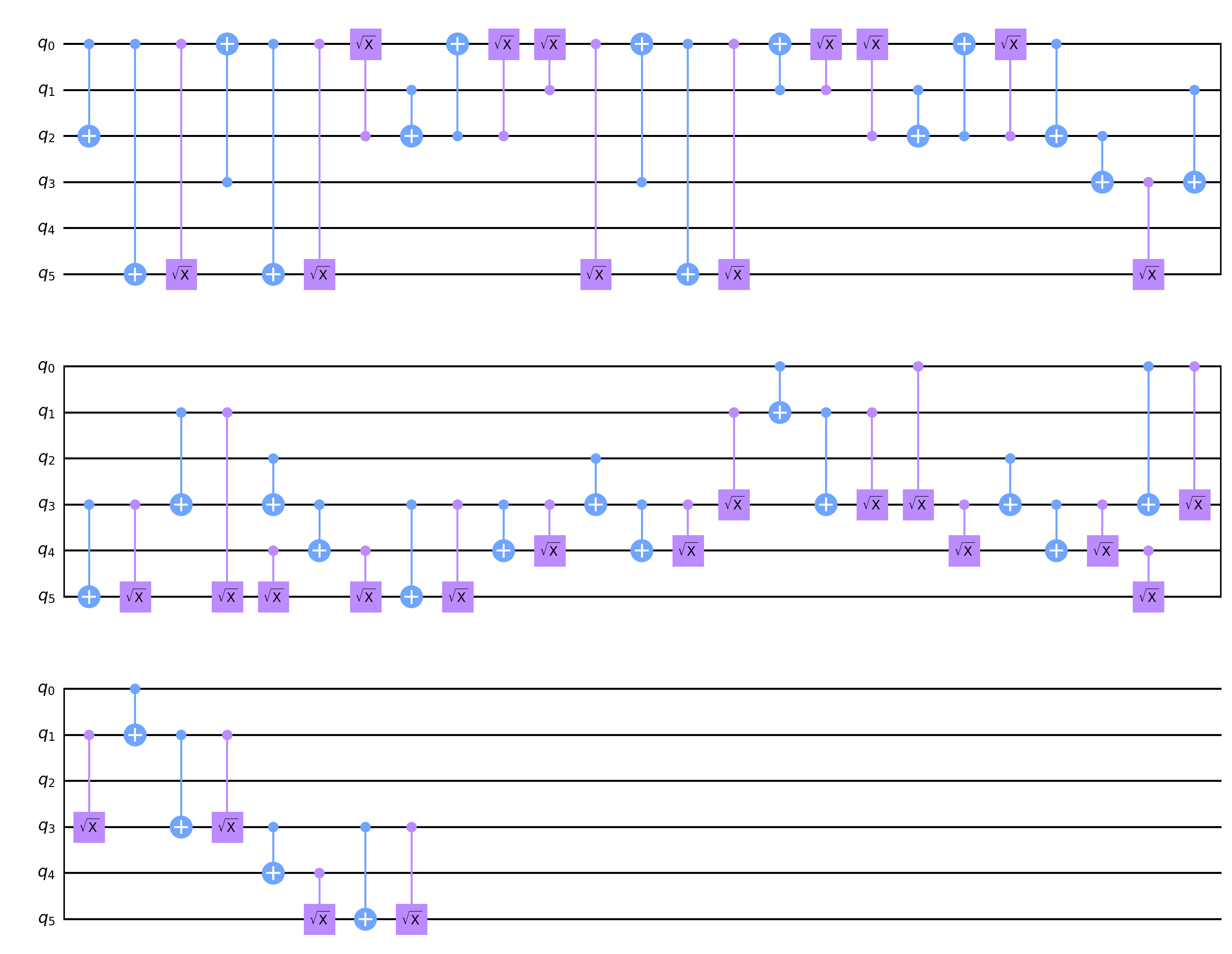}}
\caption{The $4gt4\_20$ quantum circuit with $NCV$ gates using IBM.}
\label{ncv on ibm}
\end{figure}

\end{example}

\section{Conclusion}
\label{sec:con}
This paper proposed two algorithms to map the $MCT$ circuit into the $NCV$ circuit for any Boolean function expressed in a Positive Polarity Reed-Muller $PPRM$ expansion. The first algorithm proposes a special algebraic form for the Boolean function to reduce the number of product terms after applying the following stages: factorizing the terms of function followed by applying the reorder method \cite{journals/qip/AhmedYE18}, then applying a special algebraic form that rearranges the terms of the Boolean function based on degree of term $d_{term}$, finally synthesizing $MCT$ circuit using different synthesis methods. The second algorithm converts the $MCT$ circuit into $NCV$ circuit by applying some decomposition methods followed by applying  simplification rules. The proposed algorithms produce more efficient quantum circuits with low quantum cost $QC$, on average, when compared with circuits synthesized by other methods.

\section*{Code and data availability}
All codes can be found online at the public repository :
\url{ https://github.com/Taghreed-hub/Decomposition-Algorithm}
\bibliographystyle{spmpsci}
\bibliography{references}

\end{document}